\documentclass[twoside,twocolumn,english,aps,prb]{revtex4}
\usepackage[T1]{fontenc}
\usepackage[latin1]{inputenc}

\makeatletter

\usepackage{graphicx}

\usepackage{babel}

\usepackage{babel}
\makeatother
\begin{document}

\title{Theory of non-Fermi liquid and pairing in electron-doped cuprates}

\author{Pavel Krotkov and Andrey V. Chubukov}

\affiliation{Condensed Matter Theory Center, Department of Physics, University
of Maryland, College Park, Maryland, 20742 \\
 Department of Physics, University of Wisconsin, 1150 University Ave,
Madison, WI 53706}

\begin{abstract}
We apply the spin-fermion model to study the normal state and pairing
instability in electron-doped cuprates near the antiferromagnetic
QCP. Peculiar frequency dependencies of the normal state properties
are shown to emerge from the self-consistent equations on the fermionic
and bosonic self-energies, and are in agreement with experimentally
observed ones. We argue that the pairing instability is in the $d_{x^{2}-y^{2}}$
channel, as in hole-doped cuprates, but theoretical $T_{c}$ is much
lower than in the hole-doped case. For the same hopping integrals
and the interaction strength as in hole-doped materials, we obtain
$T_{c}\sim10$K at the end point of the antiferromagnetic phase. We
argue that a strong reduction of $T_{c}$ in electron-doped cuprates
compared to hole-doped ones is due to critical role of the Fermi surface
curvature for electron-doped materials. The $d_{x^{2}-y^{2}}$-pairing
gap $\Delta(\mathbf{k},\omega)$ is strongly non-monotonic along the
Fermi surface. The position of the gap maxima, however, does not coincide
with the hot spots, as the non-monotonic $d_{x^{2}-y^{2}}$ gap persists
even at doping when the hot spots merge on the Brillouin zone diagonals.

\end{abstract}
\maketitle
\newcommand{\sgn}{\mathop{\mathrm{sgn}}\nolimits}

\newcommand{\const}{\mathop{\mathrm{const}}\nolimits}

\newcommand{\ts}{\textstyle}

\newcommand{\f}[1]{\mbox {\boldmath\(#1\)}}

\section{Introduction}

The fascinating properties of high-temperature superconductors continue
to attract high interest of condensed-matter community over the last
two decades. Most of the extensive experimental and theoretical studies
of the cuprates have been performed on hole-doped materials, such
as La$_{2-x}$Sr$_{x}$CuO$_{4}$, YBa$_{2}$Cu$_{3}$O$_{6+x}$ (YBCO),
Bi$_{2}$Sr$_{2}$CaCu$_{2}$O$_{8+x}$(Bi-2212), Tl$_{2}$Ba$_{2}$CuO$_{6+x}$,
etc. In recent years, however, there has been growing interest in
the properties of electron-doped cuprates such as Nd$_{2-x}$Ce$_{x}$CuO$_{4}$
and Pr$_{2-x}$Ce$_{x}$CuO$_{4}$~\cite{damascelli,blumberg02}.

Hole-doped and electron-doped cuprates are in many respects similar.
For both, optical conductivity measurements at small doping show a
charge-transfer gap of about $2$eV~\cite{basov,onose04}, and the
electronic Fermi surface, measured by ARPES in electron-doped cuprates
is reasonably described the same combination of hopping integrals
as in hole-doped cuprates~\cite{damascelli}. This implies that both
electron-doped and hole-doped materials are likely described by the
same underlying Hubbard-type model~\cite{millis04}. The normal state
behavior of optical conductivity near optimal doping is also very
similar in hole-doped and electron-doped materials ~\cite{marel,homes}.
Most importantly, the superconductivity in both types of materials
has $d_{x^{2}-y^{2}}$ symmetry, as evidenced by, e.g., ARPES measurements
of the momentum dependence of the pairing gap~\cite{camp,sato}.

On the other hand, the phase diagrams of hole-doped and electron-doped
cuprates are somewhat different. Electron-doped cuprates have a much
wider range of antiferromagnetism than hole-doped materials~\cite{el-pd,Aiff03}.
There is a strong evidence that the Mott physics is not very relevant
in electron-doped cuprates near optimal doping as the optical data
show~\cite{onose04} that the $1.7$eV charge-transfer gap completely
melts away as the electron doping approaches its optimal value $x\sim0.15$.
There is also no analog in electron-doped cuprates of the pseudogap
phase between the antiferromagnetic and superconducting phases on
the phase diagram. Some pseudogap behavior in the normal state has
been detected in Nd$_{1.85}$Ce$_{0.15}$CuO$_{4}$, but the onset
temperature for this behavior clearly tracks the Neel temperature
~\cite{koitzsch03}, and so a (narrow) pseudogap phase is likely
just a magnetic fluctuation regime of a quasi-2D antiferromagnet \cite{zimmers05}.
In the absence of the pseudogap, the phase diagram of electron-doped
cuprates resembles a {}``typical'' quantum-critical phase diagram~\cite{piers}
- the superconducting $T_{c}(x)$ forms a dome above the antiferromagnetic
quantum-critical point (QCP) \cite{dagan04}.

The superconducting properties of the two classes of materials also
differ. First, $T_{c}$ in electron-doped cuprates is around $10-20$K
which is almost an order of magnitude smaller than in hole-doped cuprates,
and up to a factor $500$ smaller than the pseudogap $T^{*}$, which,
as a large group of researchers believe, is the onset temperature
of the pairing without coherence in hole-doped cuprates. Second, the
pairing gap in electron-doped cuprates varies non-monotonically along
the Fermi surface -- it increases at deviations from Brillouin zone
diagonals, passes through a maximum, and then decreases. The non-monotonic
behavior of the $d-$wave gap has been originally introduced to explain
Raman data~\cite{blumberg02}. Later, the non-monotonic gap has been
directly observed in ARPES experiments ~\cite{blumberg02,sato}.
In hole-doped cuprates, the $d-$wave gap is monotonic, within error
bars~\cite{camp}.

The large difference between $T_{c}$ in electron and hole-doped cuprates
(and even larger difference with $T^{*}$ for the hole-doped materials)
despite apparently similar hopping integrals and the strength of the
Hubbard interaction calls for an explanation. In this paper, we argue
that the primary difference between the pairing instability temperature
in electron-doped and hole-doped materials is that in hole-doped cuprates
the pairing predominantly involves antinodal fermions, while for electron
doped cuprates the pairing involves fermions near zone diagonals.
We show that for the pairing of near-diagonal fermions, the Fermi
surface curvature has a very strong and negative effect on $T_{c}$,
and reduces it by more than two orders of magnitude. This eventually
leads to $T_{c}\sim10$K in electron-doped cuprates.

To solve the pairing problem, one needs to select a pairing mechanism.
Following earlier works by one of us and the others ~\cite{pines-scalapino,manske00,abanov03},
we assume that the pairing is of electron rather than phonon origin,
and that a dominant pairing interaction between fermions is mediated
by collective, Landau-overdamped magnetic fluctuations. The idea of
spin-fluctuation pairing was earlier applied to hole-doped cuprates,
with the assumption that Mott physics does not play a substantial
role in the normal (not pseudogap) state, and becomes relevant only
well inside the pseudogap phase. For hole-doped cuprates, the Luttinger
Fermi surface is such that the hot spots (Fermi surface points separated
by the antiferromagnetic momentum) are located near the corners of
the Brillouin zone. The pairing near a QCP predominantly involves
fermions from around the hot spots, and the region of pairing instability
forms a dome on top of the antiferromagnetic QCP~\cite{abanov03}.
The instability temperature increases with underdoping and saturates
at $T^{*}\sim0.03{\bar{g}}$~\cite{finn}, where ${\bar{g}}$ is
the effective interaction, which is comparable to the charge transfer
gap. Using $1.5$eV for ${\bar{g}}$ yields $T^{*}\sim500$K, which
is consistent with the onset temperature for the pseudogap behavior.
The relation between $T^{*}$ and the actual $T_{c}$ is a separate
issue which we will not discuss in this paper.

We will apply the same itinerant spin-fermion model to study the spin-mediated
pairing in electron-doped cuprates. Like we said, the electron-doped
cuprates seem even more suitable for an itinerant description than
the hole-doped materials as there is no non-magnetic pseudogap, and
the charge-transfer gap melts away near optimal electron doping. We
assume, following RPA studies of the static spin susceptibility, that
antiferromagnetism sets in at around the doping when the hot spots
merge on the Brillouin zone diagonals~\cite{manske00,markiewicz03,onufrieva04}.
At this doping the antiferromagnetic Brillouin zone touches the Fermi
surface at the diagonal points $\mathbf{k}_{F}=(\pi/2,\pi/2)$ (see
Fig. \ref{cap:sketch}), so that $2\mathbf{k}_{F}$ coincides with
the antiferromagnetic wave vector $\mathbf{Q}$. At smaller dopings,
the system is magnetically ordered, and the fermionic Fermi surface
displays electron and hole pockets.

We first consider normal state properties near a $2k_{F}$ antiferromagnetic
QCP. We reproduce and extend earlier result of Altshuler, Ioffe, and
Millis~\cite{altshuler95} that at QCP, the self-energy of a nodal
fermion has a non-Fermi liquid form and scales as $\Sigma(\omega)\propto\omega^{a}$.
The exponent $a$ is close to unity though itself varies with frequency.
In this situation, the normal state behavior is close to that in a
marginal Fermi liquid.

We next consider the pairing of these non-Fermi liquid fermions and
show that at QCP, the system is still unstable towards $d_{x^{2}-y^{2}}$
gap opening, i.e., there is a dome of the pairing instability around
critical doping where antiferromagnetism sets in. The pairing instability
temperature at QCP (which for electron-doped systems we will label
as $T_{c}$ because of the absence of the pseudogap) still scales
with ${\bar{g}}$ but because of the strong de-pairing effect associated
with the Fermi surface curvature, $T_{c}\sim0.0006{\bar{g}}$ for
the actual curvature of the cuprates Fermi surface. This is 50 times
smaller than $0.03{\bar{g}}$ for hole-doped cuprates. The uncertainty
of $T_{c}$ due to the approximate nature of the estimate of the curvature
is actually very weak as $T_{c}$ as a function of the curvature is
almost flat in a wide range of the curvatures.

We also show that the pairing gap is non-monotonic along the Fermi
surface, and passes through a maximum at some deviation from the zone
diagonal. The location of the gap maxima does not track hot spots
as at $2k_{F}$ QCP, the hot spots are right along the zone diagonals.

That an antiferromagnetically mediated $d_{x^{2}-y^{2}}$ pairing
survives in a situation when the hot spots are along the zone diagonals
is not intuitively obvious, for the strongest pairing interaction
involves quasiparticles for which the $d_{x^{2}-y^{2}}$ superconducting
gap vanishes. However, one can easily see that the separation between
the near-nodal fermions with opposite signs of the $d_{x^{2}-y^{2}}$
gap is on average closer to $\mathbf{Q}$ than that between fermions
with the same sign of the gap (see Fig. \ref{cap:sketch} b). Because
of this difference, there is still attraction in the $d_{x^{2}-y^{2}}-$wave
channel for the spin-fluctuation mediated interaction. Furthermore,
we will see that for the pairing at QCP, the kernel of the gap equation
actually does not have any extra smallness associated with the hot
spots location along the zone diagonals, i.e., $T_{c}$ is formally
of order ${\bar{g}}$. This happens because the gap equation relates
$\Delta(\mathbf{k})$ near the two diagonal hot spots $\mathbf{k}_{F}$
and $\mathbf{k}_{F}+\mathbf{Q}$ on the Fermi surface, where both
gaps are linear in the deviation from the diagonal. Then the gap equation
effectively becomes an equation on the slope of $\Delta(\mathbf{k})$
and the latter one does not contain any extra smallness, as we will
see.

The $d_{x^{2}-y^{2}}$ pairing near $2k_{F}$ instability has been
earlier considered within BCS theory~\cite{yakovenko04}, where the
authors found that $d-$wave $T_{c}$ remains small but finite at
QCP. Our results agree with Ref. \cite{yakovenko04} in that the $d-$wave
attraction survives at QCP, but we argue that the pairing involves
non-Fermi liquid fermions, and that $T_{c}$ scales with the upper
boundary of the quantum-critical regime (modulo small prefactor) rather
than with the upper boundary of the Fermi liquid behavior. The strong
reduction of $T_{c}$ in electron-doped cuprates compared to the onset
temperature for the pairing in hole-doped cuprates was also obtained
in FLEX studies of magnetically-mediated pairing~\cite{manske00}
(although the difference was less drastic than in our case). We believe
that the numerical results of Ref. \cite{manske00} describe the same
physics as our analysis. Non-monotonic variation of the $d-$wave
gap also agrees with recent numerical studies~\cite{num}.

A short summary of our results was presented earlier in \cite{krotkov06}.

The paper is organized as follows. In Section \ref{sec:Model} we
briefly review the spin-fermion approach. In Section \ref{sec:Normal-State-Properties}
we study the properties of electron-doped materials in the normal
state. In subsection \ref{sec:Perturbation-approach} we develop perturbation
approach to fermionic and bosonic self-energies. Self-consistent solution
of the Dyson's equations on the fermionic and bosonic self-energies
is covered in subsection \ref{sec:Self-consistent-self-energies}.
In subsection \ref{sec:Conductivity} we use the obtained self-energy
to find the frequency dependences of the conductivity and Raman absorption.
In Section \ref{sec:Pairing-problem} we address the pairing problem
and obtain $T_{c}$ and the momentum dependence of the gap. In \ref{sub:Solution-at-r0}
we consider the limit when the parameter representing curvature of
the Fermi surface in the gap equation formally tends to zero. In \ref{sub:Solution-at-finite}
we consider the case of finite curvature. We present numerical solution
in \ref{sub:Numerical-solution}, and an approximate analytical solution
in \ref{sub:Toy-model}, and find that the strong variation of $T_{C}$
with the curvatures can be well understood. In Section \ref{sec:Summary}
we sum up the main conclusions. Some technical details are relegated
to Appendices.

\section{Model\label{sec:Model}}

In this paper we apply the spin-fermion model to a particular form
of electron dispersion in two dimensions. In general, low energy dynamics
of strongly-interacting electrons may be studied by integrating out
the high-energy part of electron-electron interaction. Close to a
collective instability, the corresponding bosonic collective mode
tends to become gapless, and has to be included in the low-energy
model. The interaction with the near-gapless collective mode alters
the fermionic dynamics and may lead to non-Fermi-liquid behavior.
The same interaction also gives rise to the pairing.

A spin-fermion model has been previously applied to hole-doped cuprates,
and we refer the reader to earlier literature~\cite{abanov03} for
the details and the justification of the model. Here we start right
with the Hamiltonian: \begin{eqnarray}
\mathcal{H} & = & \sum_{\mathbf{k}}\epsilon_{\mathbf{k}}c_{\mathbf{k}\alpha}^{\dagger}c_{\mathbf{k}\alpha}+\sum_{\mathbf{q}}\chi_{0}^{-1}(\mathbf{q})\mathbf{S}_{\mathbf{q}}\mathbf{S}_{-\mathbf{q}}\nonumber \\
 &  & +g\sum_{\mathbf{k},\mathbf{q}}c_{\mathbf{k}+\mathbf{q},\alpha}^{\dagger}\f\sigma_{\alpha\beta}c_{\mathbf{k}\beta}\mathbf{S}_{\mathbf{q}},\label{apr-23-1}\end{eqnarray}
 Here $c_{\mathbf{k}}$ are the low-energy fermions and $\mathbf{S}_{\mathbf{q}}$
are their collective bosonic fluctuations in the spin channel. The
spin-fermion interaction is described by the effective coupling $g$;
self-consistency of the model requires $g$ to be smaller, or, at
most, of the order of the bandwidth $g\leq W$. Static spin susceptibility
$\chi_{0}(\mathbf{q})$ is determined by the high-energy physics.
We assume that it has Ornstein-Zernike form near the antiferromagnetic
vector $\mathbf{Q=}(\pi,\pi)$: \begin{equation}
\chi_{0}(\mathbf{q})=\chi_{0}/(\xi^{-2}+(\mathbf{q}-\mathbf{Q})^{2})\label{eq:chi0}\end{equation}
 The correlation length $\xi$ parameterizes proximity to the instability.
At QCP which we only study below, $\xi^{-1}=0$. The interaction $g$
and $\chi_{0}$ appear only in combination ${\bar{g}}=g^{2}\chi_{0}$
which is the actual (measurable) spin-fermion interaction.

While the static spin susceptibility is an input, the low-frequency
\emph{dynamic} part of the susceptibility (\ref{eq:chi0}) should
be determined within the model, together with the fermionic self-energy.
The reasoning is that the damping of the bosonic fluctuations occurs
only by a decay into particle-hole pairs with energies smaller than
the bosonic frequency. Because (\ref{eq:chi0}) is peaked at the antiferromagnetic
wave vector $\mathbf{Q}$, the interaction between fermions and bosons
will predominantly involve fermions near the \emph{hot spots} (Fermi
surface points separated by $\mathbf{Q}$).

The computational tool to solve the model of Eq. (\ref{apr-23-1})
at large $\xi$ is Eliashberg-type theory in which one solves for
fully renormalized fermionic and bosonic self-energies while simultaneously
neglecting vertex corrections and the self-energy dependence on momentum
transverse to the Fermi surface. This computational procedure is similar,
but not equivalent to FLEX. Eliashberg theory at strong coupling has
been discussed extensively in early literature~\cite{abanov03} and
we will just use it in our work without further discussion.

\section{Normal-State Properties\label{sec:Normal-State-Properties}}

\subsection{Perturbation approach\label{sec:Perturbation-approach}}

\begin{figure}
\includegraphics[%
  width=1\columnwidth]{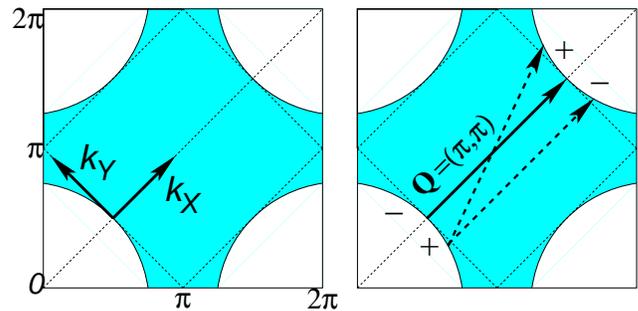}

\caption{Left: Fermi surface at the antiferromagnetic QCP with $2k_{F}=(\pi,\pi)$.
Diamond-shaped dashed lines bound the magnetic Brillouin zone. The
diagonal points of the Fermi surface (nodal points of the $d_{x^{2}-y^{2}}$-wave
gap) now become {}``hot''. Right: Graphic explanation of the attraction
in the $d_{x^{2}-y^{2}}$ channel: parts of the Fermi surface on the
same side of the zone diagonals are on average closer to the $\mathbf{Q}$
separation than the parts on the opposite sides, leading to attraction
in the $d_{x^{2}-y^{2}}$ channel (plus and minus are the signs of
the $d_{x^{2}-y^{2}}$ gap). \label{cap:sketch}}
\end{figure}

The system behavior at $2k_{F}$ QCP within Eliashberg theory has
been studied by Altshuler et al.~\cite{altshuler95}. They used RG
procedure to sum up the most divergent diagrams and obtained fermionic
and bosonic self-energies at the lowest frequencies in the normal
state. We obtained very similar results using a somewhat different
computational procedure (see below). We also obtained fermionic self-energy
at intermediate frequencies which are mostly relevant for the pairing
problem.

A set of self-consistent Dyson's equations for the fermionic and bosonic
self-energies in the normal state is\begin{eqnarray}
\chi^{-1}(\mathbf{q},\omega) & = & \chi_{0}(\mathbf{q})^{-1}+\Pi(\mathbf{q},\omega),\label{eq:Chi}\\
G^{-1}(\mathbf{k},\omega) & = & G_{0}(\mathbf{k},\omega)^{-1}+i\Sigma(\mathbf{k},\omega).\label{eq:G}\end{eqnarray}
 The Green's function of noninteracting fermions is\begin{equation}
G_{0}^{-1}(\mathbf{k},\omega)=i\omega-\epsilon_{\mathbf{k}},\label{eq:G0}\end{equation}
 and $\epsilon_{\mathbf{k}}$ is the electron dispersion. Bosonic
self-energy (polarization operator) contains contributions from non-umklapp
and umklapp scattering between hot spots separated by $\mathbf{Q}$.
Hereafter we denote by $\mathbf{q}$ the deviation of $\mathbf{q}$
from the antiferromagnetic wave vector $\mathbf{Q}$ unless otherwise
noted. In these units, the total polarization operator is $2(\Pi(\mathbf{q},\Omega)+\Pi(-\mathbf{q},\Omega))$,
where \begin{equation}
\Pi(\mathbf{q},\Omega)=2g^{2}\int\frac{d\omega d^{2}\mathbf{k}}{(2\pi)^{3}}G(\mathbf{k},\omega)G(\mathbf{k}+\mathbf{q},\omega+\Omega).\label{eq:Pi}\end{equation}
 Then \begin{equation}
\chi(\mathbf{q},\Omega)=\frac{\chi_{0}}{\mathbf{q}^{2}+2\chi_{0}\left(\Pi(\mathbf{q},\Omega)+\Pi(-\mathbf{q},\Omega)\right)}.\label{eq:ChiWithPhi0}\end{equation}
 The peculiarity of a spin-fermion theory near a $2k_{F}$ instability
is that the Fermi velocities at the hot spots are antiparallel, and
to keep $\Pi(\mathbf{q},\Omega)$ finite, one has to expand the spectrum
of the fermions in the vicinity of these points up to the second order
in the momentum component along the Fermi surface:\begin{eqnarray}
\epsilon_{\mathbf{k}} & \approx & v_{F}k_{x}+\beta^{2}k_{y}^{2},\label{eq:ep}\\
\epsilon_{\mathbf{k}+\mathbf{Q}} & \approx & -v_{F}k_{x}+\beta^{2}k_{y}^{2}.\end{eqnarray}
 Here the momentum components $k_{x}$, $k_{y}$ are normal and tangential
to the Fermi line respectively, and $\mathbf{k}$ is measured from
the hot spot (see Fig. \ref{cap:sketch}); $\beta$ parameterizes
the curvature of the Fermi line, $\kappa=2\beta^{2}/v_{F}$. The radius
of curvature used in \cite{altshuler95} is $k_{\kappa}=1/\kappa$.
The special case of a nested Fermi surface has been analyzed in ~\cite{nested}
A. Virosztek and J. Ruvalds, Phys. Rev. B \textbf{42}, 4064 (1990).

For noninteracting fermions with the Green's function (\ref{eq:G0}),
Eq. (\ref{eq:Pi}) yields\begin{equation}
\Pi_{0}(\mathbf{q},\Omega)=\frac{g^{2}}{2\pi v_{F}\beta}\sqrt{\sqrt{\Omega^{2}+E_{\mathbf{q}}^{2}}+E_{\mathbf{q}}},\label{eq:PiBare}\end{equation}
 where \begin{equation}
E_{\mathbf{q}}=-v_{F}q_{x}+\beta^{2}q_{y}^{2}/2.\label{eq:Eq}\end{equation}

Already this zeroth-order expression (\ref{eq:PiBare}) differs in
two important ways from previously studied case~\cite{abanov03}
when the two hot spots separated by $\mathbf{Q}$ had nearly orthogonal
velocities. First, at $\mathbf{q}=\mathbf{Q}$ $\Pi(\Omega)$ scales
as $\sqrt{\Omega}$ instead of the conventional $|\Omega|$, and,
moreover, diverges when $\beta\rightarrow0$ , so curvature of the
Fermi line cannot be neglected. Later we will see that $\beta\ne0$
is even more important for the pairing. It is interesting to trace
how (\ref{eq:PiBare}) converts into a conventional Landau damping
when the hot spots are moving away from the zone diagonals. This is
done in Appendix \ref{sec:Transformation-of-Pi}.

Second, a conventional Landau damping $|\Omega|$ term represents
an anomaly in analytical properties of bosonic self-energy. External
frequency $\Omega$ appears as a factor because it regularizes the
integral over internal frequencies by separating the poles into opposite
semi-planes. At the same time, the polarization operator (\ref{eq:PiBare})
gives a fractional power of external frequency not as a result of
anomaly, but simply because $\Omega$ sets an effective cutoff for
the ultraviolet-divergent integral over internal $\omega$. This distinction
becomes relevant at a finite $T$: while Landau damping is a purely
quantum phenomenon, and so does not depend on temperature, the self-energy
(\ref{eq:PiBare}) possesses $\omega/T$-scaling. Evaluating the polarization
operator for free fermions but at a finite $T$, we obtain \begin{eqnarray}
\widetilde{\Pi}_{0}(\mathbf{q},\Omega) & = & \frac{g^{2}}{2\pi v_{F}\beta}\int_{-\infty}^{\infty}\frac{\sqrt{\sqrt{4\omega^{2}+\Omega^{2}}+2\omega}}{4T\cosh^{2}\frac{\omega-E_{\mathbf{q}}/2}{2T}}d\omega\label{eq:PiIntegral}\\
 & = & \sqrt{T}f(\Omega/T,E_{\mathbf{q}}/T),\end{eqnarray}
 where\begin{equation}
f(a,b)=\frac{g^{2}}{2\pi v_{F}\beta}\int_{-\infty}^{\infty}\frac{\sqrt{\sqrt{4x^{2}+a^{2}}+2x}}{4\cosh^{2}((x-b)/2)}dx.\label{eq:PiIntegral2}\end{equation}
 At $\mathbf{q}=0$, $\Omega=0$, the polarization operator is simply:\begin{equation}
\widetilde{\Pi}_{0}=\frac{g^{2}\sqrt{\pi T}}{2\pi v_{F}\beta}|\zeta(\ts\frac{1}{2})|(\sqrt{2}-1)=\const\sqrt{T},\end{equation}
 where $\zeta(\ts\frac{1}{2})\approx-1.46$ is the Riemann Zeta function~\cite{comment}.
When $T\rightarrow0$, the function $1/4T\cosh^{2}(x/2T)$ in the
denominator of (\ref{eq:PiIntegral}) becomes a $\delta(x)$-function,
and (\ref{eq:PiIntegral}) converts to (\ref{eq:PiBare}).

The electron self-energy is given by \begin{equation}
\Sigma(\mathbf{k},\omega)=3ig^{2}\int\frac{d\omega'd^{2}\mathbf{k}'}{(2\pi)^{3}}\chi(\mathbf{k}-\mathbf{k}',\omega-\omega')G(\mathbf{k}',\omega'),\label{eq:Sigma}\end{equation}
 The coefficient $3=\f\sigma\f\sigma$ in (\ref{eq:Sigma}) comes
from the three components of the bosonic spin fluctuations. In Eliashberg
theory, the self-energy depends on frequency and on the momentum component
along the Fermi surface. We parameterize the position of the quasiparticle
on the Fermi surface by the momentum component $k_{y}$ (see Fig.
\ref{cap:sketch}), i.e., label $\Sigma(\mathbf{k},\omega)$ as $\Sigma(k_{y},\omega)$
along the Fermi surface. This does not imply that $k_{x}$ is zero
as $k_{x}$ and $k_{y}$ are related: from $\epsilon_{k}=0$, $k_{x}=-\beta^{2}k_{y}^{2}/v_{F}$
along the Fermi surface. Since fermions are fast compared to bosons,
the integral over momenta in (\ref{eq:Sigma}) can be factorized:
the one perpendicular to the Fermi surface involves only fast electrons,
and the one along the Fermi surface ($\epsilon_{\mathbf{k}'}=0$)
involves slow bosons. The corrections from keeping the momentum transverse
to the Fermi surface in the bosonic propagator is of the same smallness
as the vertex correction (in this, Eliashberg theory differs from
FLEX where the momentum integration is not factorized)

Substituting \begin{equation}
G(\mathbf{k},\omega)^{-1}=i(\omega+\Sigma(\mathbf{k},\omega))-\epsilon_{\mathbf{k}}\label{eq:Gfull}\end{equation}
 into (\ref{eq:Sigma}) and integrating over momenta transverse to
the Fermi surface, we obtain for the self-energy \begin{equation}
\Sigma(k_{y},\omega)=\int_{-\omega}^{\omega}\frac{d\Omega}{2\pi}\int dk'_{y}\widetilde{\chi}(k,k'_{y},\Omega),\label{eq:SigmaT0}\end{equation}
 The reduced bosonic propagator \begin{equation}
\widetilde{\chi}(k_{y},k'_{y},\Omega)=\frac{3g^{2}}{4\pi v_{F}}\left.\chi(\mathbf{k}'-\mathbf{k},\Omega)\right|_{\epsilon_{\mathbf{k}}=\epsilon_{\mathbf{k}'}=0}\label{eq:ChiTilde}\end{equation}
 is taken between the two points $\mathbf{k}$ and $\mathbf{k}'$
on opposite sheets of the Fermi surface.

The reduced propagator depends on the curvature in two ways. First,
the polarization operator $\Pi$ depends on curvature via the overall
factor and the $E_{\mathbf{q}}$ term in Eq. (\ref{eq:Eq}). Second,
there is a direct $\beta$ dependence in the static part of the susceptibility,
as $(k_{x}-k_{x}')^{2}$ term in $\chi$ reduces to $\beta^{4}(k_{y}^{2}+k'_{y}{}^{2})^{2}/v_{F}^{2}$
once we use the Fermi surface relations $k_{x}=-\beta^{2}k_{y}^{2}/v_{F}$
and $k'_{x}=\beta^{2}k'_{y}{}^{2}/v_{F}$.

For the self-energy exactly at the hot spot $(\pi/2,\pi/2)$, the
first dependence is much more relevant than the second one. Indeed,
setting the external momentum $k=0$ and taking $k'$ to be at the
Fermi surface we find that $E_{\mathbf{q}}$ and $E_{-\mathbf{q}}$
terms with $\mathbf{q}=\mathbf{k}'$ reduce to $E_{\mathbf{k}'}=-\beta^{2}k_{y}^{\prime}{}^{2}/2$,
and $E_{-\mathbf{k}'}=3\beta^{2}k_{y}^{\prime}{}^{2}/2$, i.e., $E_{\pm\mathbf{k}'}$
scales with the curvature. The polarization operators $\Pi(\mathbf{q},\Omega)$
and $\Pi(-\mathbf{q},\Omega)$ given by (\ref{eq:PiBare}) depend
on the ratio $\Omega/E_{\pm\mathbf{k}'}$, and therefore the dependence
of $E_{\pm\mathbf{k}'}$ on the curvature is relevant. On the other
hand, the $\beta^{4}k'_{y}{}^{4}/v_{F}^{2}$ term in $\chi_{0}^{-1}$
is clearly subleading to $k'_{y}{}^{2}$ at small $k_{y}$.

Keeping the dependence on the curvature only in the polarization operator
we obtain that away from the QCP, when $\xi^{-1}\ne0$, $\Sigma(k_{y}=0,\omega)=\Sigma(\omega)$
has a usual Fermi liquid expansion in powers of $\omega$. At the
QCP the Fermi liquid behavior breaks down, and $\Sigma$ acquires
non-Fermi-liquid frequency dependence. To find it, we assume and then
verify that typical $k'_{y}$ in the integral for the self-energy
are much larger than typical $\Omega$, and expand the denominator
in (\ref{eq:ChiWithPhi0}) in frequency. Using $E_{\mathbf{q}}=-\beta^{2}k'_{y}{}^{2}/2$
and $E_{-\mathbf{q}}=3\beta^{2}k'_{y}{}^{2}/2$ valid when ${\mathbf{k}'}$
is at the Fermi surface, we find that the expansion holds in $k'_{y}{}^{2}+\sqrt{3}q_{0}|k'_{y}|+|\Omega|q_{0}^{3}/4\omega_{0}|k'_{y}|$,
where \begin{eqnarray}
q_{0} & = & \bar{g}/\pi v_{F},\label{eq:q0}\\
\omega_{0} & = & (\bar{g}\beta/2\pi v_{F})^{2}.\label{eq:omega0}\end{eqnarray}

The extra static term $\sqrt{3}q_{0}|k'_{y}|$ coming from $\Pi$
just reflects the fact that once $k'$ is nonzero, the actual distance
between the diagonal Fermi surface point and a Fermi surface point
on the opposite sheet of the Fermi surface is not exactly $(\pi,\pi)$,
so the susceptibility acquires an additional {}``mass'' term. Note
also that the leading term in the frequency expansion has a conventional
$|\Omega|/|k'_{y}|$ term, typical for small momentum scattering.

Substituting this expansion into ${\tilde{\chi}}(q=k'_{y},\Omega)$
and integrating over $k'_{y}$ in (\ref{eq:SigmaT0}), we obtain that
the momentum integral is logarithmic and is cut from below by frequency
$\Omega$ (this justifies the assumption that typical momenta are
larger than $\Omega$). Integrating finally over frequency, we obtain
\begin{equation}
\Sigma(\omega)=-\frac{\sqrt{3}}{4\pi}\omega\log\frac{|\omega|}{\omega_{0}}.\label{eq:iSigma}\end{equation}
 This marginal Fermi liquid, $\omega\log\omega$, behavior of the
self-energy at small frequencies was first detected in~\cite{altshuler95}.
A simple analysis shows that it extends to frequencies of order $\omega_{0}$.
At larger frequencies $\omega>\omega_{0}$, typical $k'_{y}$ in the
integral for the self-energy are of order $q_{0}(\Omega/\omega_{0})^{1/4}$.
At these momenta, $E_{\pm\mathbf{q}}\sim(\Omega\omega_{0})^{1/2}<\Omega$,
so the polarization operator can well be approximated by its zero
momentum form $\Pi(0,\Omega)\propto\sqrt{\Omega}$. Substituting this
form into (\ref{eq:SigmaT0}), we find \begin{equation}
\Sigma(\omega)=\omega_{0}^{1/4}|\omega|^{3/4}\sgn\omega.\label{eq:w3/4}\end{equation}

At finite $k_{y}$, the form of the self-energy $\Sigma(k_{y},\omega)$
is rather involved. Below we will only need self-energy at $\omega>\omega_{0}$
since in electron-doped cuprates $\omega_{0}\sim10$meV is small (see
below). Parametrically, the leading dependence comes from the polarization
operator, via $v_{F}q_{x}$ term in $E_{\mathbf{q}}$, which for the
particles at the Fermi surface reduces to $\beta^{2}(k_{y}^{2}+k'_{y}{}^{2})\sim\beta^{2}k_{y}^{2}$
as $(k_{y}-k'_{y})^{2}$ is small and irrelevant. Substituting this
into $\Pi(q,\Omega)$ and evaluating the self-energy, we obtain \begin{equation}
\Sigma(k_{y},\omega)=\omega_{0}^{1/4}|\omega|^{3/4}\sgn\omega~\frac{1}{\left(1+a\left(\frac{\beta^{2}k_{y}^{2}}{\omega}\right)^{2}\right)^{1/8}}.\label{eq:w3/4-1}\end{equation}
 where $a=O(1)$. This formula shows that the self-energy begins decreasing
at deviations from the nodal direction $k_{y}^{2}\sim\Omega/\beta^{2}$.
At the same time, the momentum dependence in (\ref{eq:w3/4-1}) is
very weak from practical point of view. The $k_{y}{}^{4}$ term in
the static part of ${\tilde{\chi}}^{-1}$ leads to a stronger $1/(1+(k_{y}/k_{0})^{4})^{1/2}$
dependence, but this happens at larger $k_{y}$ as $k_{0}\propto\omega^{1/8}$.
We assumed that the self-energy scaling behavior is roughly in between
the two analytical dependences -- the self-energy $\Sigma(k_{y},\omega)$
rather weakly depends on $k_{y}$ up to some $k_{y}$ which approximately
scales as $\omega^{1/4}$, and smoothly decreases at larger $k_{y}$.

Finally, for the calculations of $T_{c}$ in Section \ref{sec:Pairing-problem},
we will need the self-energy at a finite $T$. We evaluated it numerically
by replacing integration over frequency by summation over discrete
Matsubara frequencies. \begin{equation}
\Sigma(k_{y},\omega_{m})=\sgn\omega_{m}T\sum_{|\Omega_{n}|<\omega_{m}}\int dk'_{y}\tilde{\chi}(k_{y},k'_{y},\Omega_{n}),\label{eq:SigmaTne0}\end{equation}
 where $\tilde{\chi}$ is the same as in (\ref{eq:ChiTilde}). Numerically
we found that the explicit temperature dependence of $\Sigma(k_{y},\omega_{m})$
is again weak at $\omega>\omega_{0}$, so the self-energy at finite
temperature has to a good accuracy the same functional form as at
$T=0$, but in discrete Matsubara frequencies.

\subsection{Self-consistent self-energies\label{sec:Self-consistent-self-energies}}

The marginal Fermi liquid form of the self-energy at $\omega<\omega_{0}$
and the $\omega^{3/4}$ dependence at $\omega>\omega_{0}$ were obtained
using the bare fermionic propagators. We need to verify whether the
results survive when calculations of the fermionic and bosonic self-energies
are performed self-consistently, using the full propagators.

Landau damping linear-in-$\Omega$ term in the polarization operator
is an anomaly, and it does not depend on whether the polarization
bubble is evaluated with bare or full fermionic propagators, as long
as the self-energy depends only on frequency. However, as we already
discussed, the $\Omega^{1/2}$ term in $\Pi$ is not an anomaly, and
the form of $\Pi(\mathbf{q}=0,\Omega)$ does generally depend on the
fermionic self-energy, which in turn depends on the functional form
of $\Pi$. This implies that lowest-order results are not sufficient,
and one has to carry out full self-consistent calculations. This was
first noticed in ~\cite{altshuler95}.

Fortunately, these calculations are not necessary at $\omega>\omega_{0}$
as at these frequencies the self-energy $\Sigma\propto\omega(\omega_{0}/\omega)^{1/4}$
is smaller than $\omega$, and the renormalized fermionic Green's
function is close to the bare one. In this situation, the expression
$\Pi(\mathbf{q},\Omega)\propto\Omega^{1/2}$ that was obtained using
free fermions is a good approximation for interacting fermions as
well, and no corrections are necessary. Next, using the fact that
the self-energy $\Sigma(\omega)$ is expressed via the fermionic density
of states, and the latter does not depend on the fermionic self-energy
(again as long as $\Sigma$ depends only on $\omega$), we find that
Eq. (\ref{eq:w3/4}) survives. Vertex corrections and other corrections
beyond Eliashberg theory can be treated in the same way as for other
quantum-critical problems~\cite{abanov03,aim,pepin}, and are irrelevant.

On the other hand, the marginal Fermi liquid form of the self-energy
at frequencies smaller than $\omega_{0}$, Eq. (\ref{eq:iSigma}),
does not survive in the higher-order corrections. In cuprates $\omega_{0}$
is small, and so this is not very relevant to our study, but it is
interesting to address it from the principle point of view and we
briefly discuss it. Altshuler et al ~\cite{altshuler95} found that
next order term yields $\Sigma_{2}(\omega)\propto\omega\log^{2}\omega$
and conjectured that higher-order logarithmic corrections form geometric
series that exponentiates to a power law \begin{equation}
\Sigma(\omega)=\omega_{0}^{\alpha}|\omega|^{1-\alpha}\sgn\omega.\label{eq:w2/3}\end{equation}
 They found $\alpha=\sqrt{3}/4\pi\approx0.14$. We obtained a nearly
identical result using a different computational procedure with does
not require that the series is geometric. Namely, we obtained a self-consistent
solution for the self-energy (\ref{eq:Sigma}) by evaluating the polarization
operator $\Pi$ (\ref{eq:Pi}) with the full Green;s function (\ref{eq:Gfull}).
We assumed that $\Sigma(\omega)\propto|\omega|^{1-\alpha}$, where
$\alpha$ is unknown, re-evaluated the polarization bubble with the
full fermionic $G$ {[}this gives $\Pi(\mathbf{q},0)\propto E_{\mathbf{q}}^{(2-\alpha)/2\alpha}{}$,
$\Pi(0,\Omega)\propto\Omega^{(2-\alpha)/2}${]}, substituted the result
into the integral for $\Sigma(\omega)$, and solved the self-consistent
equation on $\alpha$. The calculations, which are detailed in Appendix
\ref{sec:Self-energy-at-T=3D3D3D0}, yield $\alpha\approx0.15$ which
is surprisingly close (although not identical) to $\alpha=0.14$ obtained
by Altshuler et al~\cite{altshuler95}.

\subsection{Conductivity and Raman Response\label{sec:Conductivity}}

We now use the result for the self-energy to compute optical conductivity
and Raman response in the normal state. Complex conductivity $\sigma(\omega)$
can be computed from the Kubo formula:\begin{equation}
\sigma_{\alpha\beta}(i\omega_{n})=\frac{1}{\omega_{n}}\left[\Pi_{\alpha\beta}^{(j)}(i\omega_{n})-\Pi_{\alpha\beta}^{(j)}(i0)\right],\end{equation}
 where $\Pi_{\alpha\beta}^{(j)}$ is the current-current correlator
with zero transmitted momentum. Dependence on real $\omega$ is found
by the transformation $i\omega_{n}\rightarrow\omega+i0$ from Matsubara
frequencies $\omega_{n}$. Assuming constant velocities $v_{F}$,
the current-current correlator $\Pi_{\alpha\beta}^{(j)}(i\omega_{n})=e^{2}v_{F}^{2}\delta_{\alpha\beta}\Pi(i\omega_{n})/2g^{2}$
is proportional to the Matsubara bubble $\Pi(i\omega_{n})$ with zero
transmitted momentum. Neglecting corrections to the current vertices
(which is justified when $\Sigma$ predominantly depends on $\omega$),
we obtain \begin{equation}
\sigma(i\omega)\propto\frac{1}{\omega_{m}}\int dk_{y}\int_{0}^{\omega_{m}}\frac{d\epsilon}{\omega_{m}+\Sigma(k_{y},\omega_{m}-\epsilon)+\Sigma(k_{y},\epsilon)}.\label{eq:sM}\end{equation}
 At zero temperature and finite frequencies which we consider, the
short-circuiting of the conductivity considered in \cite{hlubina95}
is not important, and the conductivity can be approximated by expanding
the denominator in (\ref{eq:sM}) in the self-energy, integrating
the self-energy over $k_{y}$, and putting the result back into the
denominator. The largest contribution to conductivity then comes from
the nodal regions where the self-energy has a non-Fermi liquid form.

To integrate the self-energy explicitly over $k_{y}$ requires substantial
computational efforts as there are two distinct sources for $k_{y}$
dependence in $\Sigma$ (see previous subsection). We carry out approximate
calculations -- we assume that $\Sigma(k_{y},\epsilon)$ is independent
on $k_{y}$ and equal to $\Sigma(k_{y}=0,\epsilon)=\Sigma(\epsilon)$
for $|k_{y}|<k_{y,\mathrm{max}}$ and then falls off rapidly, such
that:\begin{equation}
\sigma(i\omega)\propto\frac{k_{y,\mathrm{max}}(\omega_{m})}{\omega_{m}}\int_{0}^{\omega_{m}}\frac{d\epsilon}{\omega_{m}+\Sigma(\omega_{m}-\epsilon)+\Sigma(\epsilon)}.\label{eq:s}\end{equation}
 Following the discussion in preceeding subsection, we assume that
for $\omega_{m}>\omega_{0}$, the threshold momentum $k_{y,\mathrm{max}}(\omega_{m})\sim\omega_{m}^{1/4}$.
We also assume by continuity that for $\omega<\omega_{0}$, $k_{y,\mathrm{max}}(\omega_{m})\sim\omega_{m}^{\alpha}$.
Since $k_{y,\mathrm{max}}$ depends on frequency, there is an ambiguity
whether we should choose the external $\omega$ or internal $\epsilon$
for the cutoff. One also should be careful when transforming from
Matsubara frequencies to the real ones. We comment on this matter
in more detail in Appendix \ref{sec:Choice-of-cutoff}. The conclusion
is basically that this ambiguity is irrelevant for the frequency dependence
of conductivity.

\begin{figure}
\includegraphics[%
  width=0.99\columnwidth]{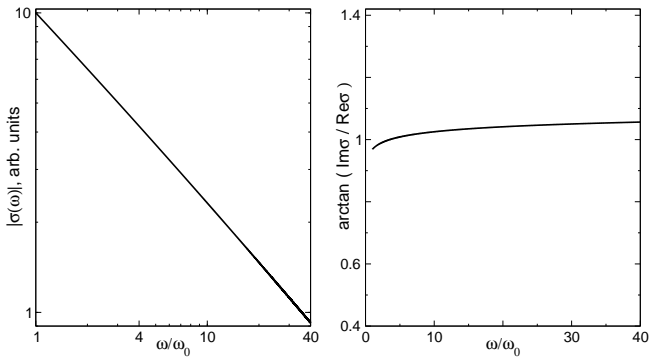}

\caption{Normal state conductivity as a function of frequency shows a scaling
behavior for $\omega_{0}<\omega<40\omega_{0}$. Left: $|\sigma(\omega)|$,
the behavior is indistinguishable from the $\sigma(\omega)\sim\omega^{-0.64}$
dependence. Right: $\arctan\Im\sigma(\omega)/\Re\sigma(\omega)$,
an almost constant value means that both $\Im\sigma(\omega)$ and
$\Re\sigma(\omega)$ scale as $\omega^{-0.64}$. \label{cap:conductivity}}
\end{figure}

At very small frequencies, $\omega<\omega_{0}$, the self-energy dominates
fermionic dynamics $\Sigma(\omega)>\omega$, and both $\Im\sigma$
and $\Re\sigma$ have the same power-law behavior: $\sigma(\omega)\propto\omega^{-1+2\alpha}\sim\omega^{-0.7}$.
At larger frequencies $\Sigma(\omega)<\omega$, and one should not
generally expect $\Re\sigma$ and $\Im\sigma$ to scale with each
other. Surprisingly, the scaling behavior extends, with almost the
same exponent -- we found that $\sigma(\omega)\propto\omega^{-0.64}$
over a very wide range up to $\sim40\omega_{0}$ (see Fig. \ref{cap:conductivity}).
Such power-law behavior is not indicative of quantum-critical scaling,
but rather a consequence of the flattening of the fermionic self-energy
at high frequencies\cite{norman06}, i.e. that $\Sigma(\omega)/\omega\propto(\omega_{0}/\omega)^{1/4}$
is a slow decaying function. The $\omega^{-\gamma}$ behavior of conductivity
with $\gamma\approx0.68$ has been observed in Pr$_{1.85}$Ce$_{0.15}$CuO$_{4}$
below $400$meV ~\cite{homes}. Both the exponent and the experimental
frequency range are quite consistent with our results. A very similar
behavior of the infrared conductivity at intermediate energies, also
caused by the flattening of $\Sigma$, has been discussed for hole-doped
cuprates \cite{norman06}.

Raman absorption measures the imaginary part of the fully renormalized
particle-hole susceptibility at vanishingly small incoming momentum,
weighted with Raman form factors that depend on the scattering geometry.
In the $B_{1g}$ geometry the Raman form factors are $\gamma_{B_{1g}}(\mathbf{k})\propto(\cos k_{x}-\cos k_{y})$.
They are the largest for fermionic momenta near $(0,\pi)$ and symmetry
related points. In the $B_{2g}$ geometry, the Raman form factors
are $\gamma_{B_{2g}}(\mathbf{k)}\propto\sin k_{x}\sin k_{y}$, and
the Raman signal comes from around $(\pi/2,\pi/2)$ points which are
close to the diagonal hot spots on the Fermi surface. For electron-doped
cuprates these are the regions where we found non-Fermi liquid behavior
of the fermionic self-energy.

Unlike the current vertex for conductivity, the Raman vertex is renormalized
by the interaction even when $\Sigma(\omega)$ depends only on frequency.
The renormalization of the $B_{1g}$ Raman vertex is not that relevant
because the interaction peaked at or near $\mathbf{Q}=(\pi,\pi)$
connects regions of the Brillouin zone where the bare vertex changes
sign. The ladder series pertaining to vertex renormalization is then
alternate in sign and roughly renormalizes $\gamma_{B_{1g}}$ to $\gamma_{B_{1g}}/(1+A\gamma_{B_{1g}})$,
where $A$ is positive. On the other hand, the $B_{2g}$ Raman vertex
$\gamma_{B_{2g}}(\mathbf{k})$ does not change sign under $\mathbf{k}\rightarrow\mathbf{k}+\mathbf{Q}$.
To a good approximation, we can approximate this vertex by a constant
$\gamma_{B_{2g}}$. The renormalization of the Raman vertex then coincides
with that of the density vertex, and like a density vertex, is then
related to the self-energy by the Ward identity: the full $\gamma_{B_{2g}}^{\mathrm{full}}(\omega)=\gamma_{B_{2g}}(1+\partial_{\omega}\Sigma(\omega))$.
The vertex then leads to the additional frequency dependence $\gamma_{B_{2g}}(\omega)\sim\partial_{\omega}\Sigma(\omega)$
such that $\gamma_{B_{2g}}(\omega)\propto\omega^{-\alpha}$ for $\omega<\omega_{0}$
and $\omega^{-1/4}$ for $\omega>\omega_{0}$.

Convoluting two Raman vertices with the polarization bubble we obtain,
approximately \begin{equation}
R_{B_{2g}}(\omega)\sim k_{y,\mathrm{max}}(\omega)\left(1+\partial_{\omega}\Sigma(\omega)\right)^{2}\frac{\omega}{\omega+\Sigma(\omega)}\end{equation}
 Substituting the forms of the self-energy and $k_{y,\mathrm{max}}(\omega)\sim\omega^{\alpha}$
for $\omega<\omega_{0}$, $k_{y,\mathrm{max}}(\omega)\sim\omega^{1/4}$
for $\omega>\omega_{0}$, we find that at $T=0$, $R_{B_{2g}}(\omega)$
is flat: it is a constant at $\omega<\omega_{0}$, and slowly crosses
over to the $(\omega/\omega_{0})^{-1/4}$ behavior at $\omega>\omega_{0}$.

The almost flat form of the Raman intensity at frequencies $\omega>\omega_{0}$
is consistent with the experimental data~\cite{quazilbash05}. The
experimental behavior at small frequencies is not flat, but according
to \cite{quazilbash05} it is dominated by temperature effects which
we do not consider here.

\section{Pairing problem \label{sec:Pairing-problem}}

\begin{figure}
\includegraphics[%
  width=0.6\columnwidth]{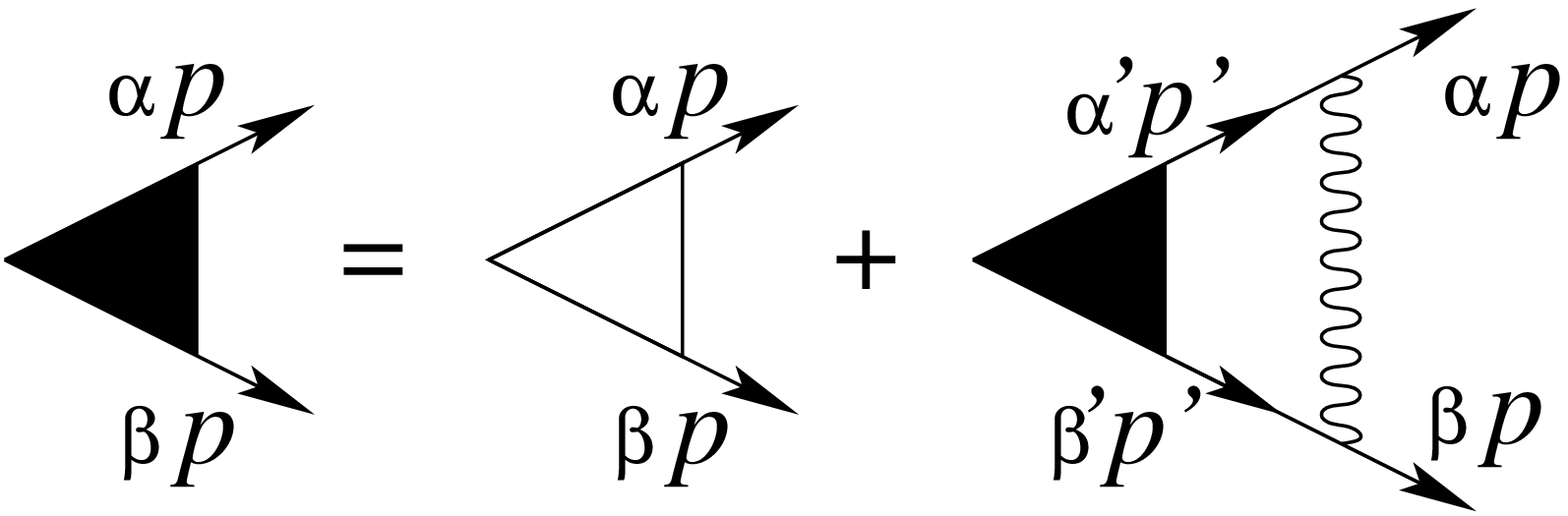}

\caption{Diagrammatic equation on the anomalous vertex \label{cap:Diagrammatic-equation-on}}
\end{figure}

The linearized gap equation at QCP is obtained using Eliashberg technique
for collective-mode mediated pairing. Assuming that the pairing occurs
in the singlet channel, we write the pairing vertex as $\Phi_{\alpha\beta}(p)=i\sigma_{\alpha\beta}^{y}\Phi(p)$
and obtain (see Fig. \ref{cap:Diagrammatic-equation-on}) \begin{equation}
\Phi(k)=-3g^{2}\sum_{k'}\Phi(k')G(k')G(-k')\chi(k-k')\label{eq:PhiSigma}\end{equation}
 where $k$ stands for a 3-component vector \textbf{$(\mathbf{k},\omega)$},
$\sum_{p}=T\sum_{\omega}\int d^{2}k/(2\pi)^{2}$, the Green's functions
are full: $G^{-1}(k)=G_{0}^{-1}+\Sigma(k)$, and the prefactor $-3$
comes from the convolutions of the Pauli matrices $\sigma_{\alpha'\beta'}^{y}\f\sigma_{\alpha\alpha'}\f\sigma_{\beta\beta'}=-3\sigma_{\alpha\beta}^{y}$.
The pairing vertex $\Phi(k)$ is related to pairing gap $\Delta(k)$
as $\Phi(\mathbf{k},\omega)=\Delta(\mathbf{k},\omega)\omega/(\omega+\Sigma(\mathbf{k},\omega))$.
The solution of this linearized gap equation yields both $T_{c}$
and the momentum and frequency dependence of $\Phi(\mathbf{k},\omega)$.
At least close to $T_{c}$, this momentum dependence must be close
to that of the true pairing gap.

As in the normal state analysis, we use the fact that bosons are slower
than fermions and approximate $\chi(k-k^{\prime})$ by Fermi surface
to Fermi surface interaction, i.e, put both $\mathbf{k}$ and $\mathbf{k}^{\prime}$
on the Fermi surface. Like before, we also assume that the fermionic
self-energy does not depend on the momentum component transverse to
the Fermi surface. We will, however, keep the momentum dependence
of $\Sigma$ along the Fermi surface. We will see below that $T_{c}$
very strongly and non-analytically depends on the Fermi surface curvature
$\kappa=2\beta^{2}/v_{F}$. This strong non-analytic dependence comes
from the dependence on curvature of the Fermi surface to Fermi surface
interaction $\chi(k-k^{\prime})$ (see below). Accordingly, we will
keep the dependence on the curvature in $\chi(k-k^{\prime})$ and
in the fermionic self-energy (because $\Sigma$ is an integral of
$\chi(k-k^{\prime})$), but neglect it in the Jacobian for the transformation
from the integration over $d^{2}k$ to the integration over $d\epsilon_{k}$,
i.e., set $dk_{x}dk_{y}=(1/v_{F})d\epsilon_{k}dk_{y}$ (the notations
are the same as in Fig. 1). Under these assumptions, the momentum
integration transverse to the Fermi surface involves the product of
two fermionic propagators and reduces to \begin{equation}
\int G(p)G(-p)d\epsilon_{\mathbf{k}}=\frac{\pi}{|\omega+\Sigma(k_{y},\omega)|}\end{equation}
 Substituting this into (\ref{eq:PhiSigma}) and using the fact that
the interaction involves momentum transfers near $Q$, we obtain \begin{equation}
\Phi_{0}(k_{y},\omega_{n})=-T\sum_{\omega_{m}}\int dk_{y}'K_{nm}(k_{y},k_{y}')\Phi_{Q}(k_{y}',\omega_{m}),\label{eq:inteq-1}\end{equation}
 where the kernel \begin{equation}
K_{nm}(k,k')=\frac{\widetilde{\chi}(k,k',\omega_{n}-\omega_{m})}{|\omega_{m}+\Sigma(k',\omega_{m})|}.\label{10-2}\end{equation}
 The reduced $\widetilde{\chi}$ is defined in (\ref{eq:ChiTilde}),
$k_{y}$ and $k_{y}^{\prime}$ are small, and the notations $\Phi_{0}$
and $\Phi_{Q}$ imply that the pairing vertices on the right and left
hand sides of (\ref{eq:inteq-1}) belong to the opposite hot spots.

Because the kernel (\ref{10-2}) is invariant under the sign inversions
of frequency and momenta, the solutions are symmetric or antisymmetric
in both. A solution with $d_{x^{2}-y^{2}}$ symmetry is antisymmetric
in $k_{y}$. It is also symmetric in frequency as at $T_{c}$ the
largest eigenvalue of the kernel reaches unity, and from a known theorem
the eigenfunction of the largest eigenvalue is symmetric in $\omega$~\cite{comm4}.
As a further simplification, we consider a {}``antiferromagnetic''
$d_{x^{2}-y^{2}}$ order parameter made out of $n=1$, $n=3$, etc.
partial components of the $B_{1g}$ representation of the $D_{4h}$
group for tetragonal symmetry (i.e., $\cos k_{x}-\cos k_{y}$, $~\cos3k_{x}-\cos3k_{y}$,
\ldots{}). For the odd $n$ solutions, the pairing vertex (and the
gap) change sign when the momentum is shifted by $Q$, i.e., $\Phi_{0}(k_{y},\omega_{n})=-\Phi_{Q}(k_{y},\omega_{n})$.
Note that a {}``conventional'' $\cos k_{x}-\cos k_{y}$ solution
falls into this category. Using this condition and dropping the subscript
we get from (\ref{eq:inteq-1}):\begin{equation}
\Phi(k_{y},\omega_{n})=T\sum_{\omega_{m}}\int dk_{y}'K_{nm}(k_{y},k_{y}')\Phi(k_{y}',\omega_{m}).\label{eq:inteq-2}\end{equation}

The remaining input is the polarization operator which is a part of
$\tilde{\chi}$. We verified \emph{a posteriori} that the dominant
contribution to the pairing comes from $\omega>\omega_{0}$, where
$\Pi(q,\Omega)$ scales as $\Omega^{1/2}$ and weakly depends on $q$.
In principle, one should use the full, finite $T$ form of of the
polarization bubble. However, the explicit temperature dependence
of $\Pi$ only complicates the calculations but does not lead to any
new physics. We assume without further discussion that the $T$ dependence
of the polarization bubble does not alter $T_{c}$ in any substantial
way and will use the same functional form $\Pi(\Omega_{m})=(g^{2}/2\pi v_{F}\beta)\sqrt{|\Omega_{m}|}$
as at $T=0$, but use discrete Matsubara frequencies at a finite $T$.
For accuracy, in numerical calculations we kept the full $T$ dependence
of the fermionic self-energy. We found, however, that the full form
of $\Sigma(k_{y},\omega_{m},T)$ differs very little from the zero-temperature
form $\Sigma(k_{y},\omega_{m})$ taken at discrete Matsubara frequencies

\begin{figure}
\includegraphics[%
  width=0.99\columnwidth]{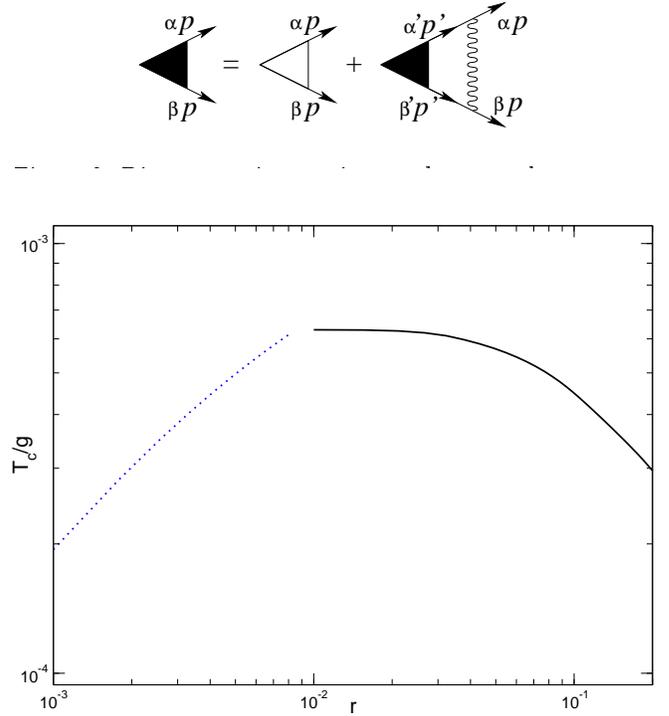}

\caption{Pairing instability temperature $T_{c}$ in units of the coupling
constant $\overline{g}$ vs. the curvature parameter $r=\bar{g}\beta^{2}/\pi v_{F}^{2}$.
Solid line gives the result of the numeric solution, which requires
increasingly long computational time as $r\rightarrow0$. Dashed line
is the analytic small-$r$ form $T_{c}=0.95\bar{g}r/(1+5r^{2/5})^{4}$.
\label{cap:Tc}}
\end{figure}

We stick to $N=2$ in this section. It is convenient to measure frequency
and temperature in units of $\omega_{0}$ (\ref{eq:omega0}) as the
fermionic self-energy at $k_{y}=0$ is (\ref{eq:w3/4}), and measure
momenta $k_{y}$ in units of $q_{0}$ (\ref{eq:q0}). Switching to
these new units, we reproduce Eqs. (\ref{eq:inteq-1}) and (\ref{eq:ChiTilde})
with \begin{equation}
\widetilde{\chi}(k,k',\Omega)=\frac{3}{2N}\frac{1}{(k-k')^{2}+r^{2}(k^{2}+k'^{2})^{2}+\sqrt{|\Omega|}}.\label{eq:ChiTildeDimless}\end{equation}
 The dimensionless quantity \begin{equation}
r=\bar{g}\beta^{2}/\pi v_{F}^{2}=\left(\frac{\bar{g}}{2\pi v_{F}p_{F}}\right)\kappa p_{F}\label{10-6}\end{equation}
 is proportional to both the curvature and the interaction. Both $k^{2}$
and $k^{4}$ terms in $\widetilde{\chi}(k,k',\Omega)$ come from the
original $(\mathbf{k}-\mathbf{k}')^{2}$ term in $\chi(k-k^{\prime})$
when we use that for the two fermions on the opposite sheets of the
Fermi surface $k_{x}=\beta^{2}k_{y}^{2}/v_{F}$ and $k'_{x}=-\beta^{2}k'_{y}{}^{2}/v_{F}$.
The normalization scale $\omega_{0}$ also scales with $r$: \begin{equation}
\omega_{0}\equiv\frac{\overline{g}}{4\pi}r\end{equation}
 and hence the curvature is present both in the overall factor, \textit{and}
in the pairing susceptibility (and in the fermionic self-energy at
a non-zero $k_{y}$, by virtue of the $r-$dependence of the susceptibility).

The presence of the curvature both in the overall normalization factor
for $T$ and in the pairing susceptibility is what \textit{qualitatively}
distinguishes the $d-$wave pairing of near-nodal fermions from the
pairing of antinodal fermions, which was previously studied in the
context of pairing in hole-doped cuprates. In the latter case, the
pairing instability temperature is finite already at zero curvature,
and the effect of the curvature on $T_{c}$ is likely small (although
this issue has not been analyzed in detail yet). In the present case,
$T_{c}$ just vanishes without curvature, so the curvature, parameterized
by $r$, plays the major role. The same $r$ also specifies momentum
dependence of $\Phi(k_{y},\omega_{n})$, and, hence, of the pairing
gap.

Below we will solve the gap equation separately for the cases $r\rightarrow0$,
when only the $r-$dependence of the overall factor matters, and for
finite $r$, when the $rk^{4}$ term in the pairing susceptibility
has to be taken into consideration. We will see that the $r-$dependence
in the denominator of (\ref{eq:ChiTildeDimless}) becomes relevant
already at very small $r$, and for $r\sim0.1$, relevant to the cuprates
(see below), the actual $T_{c}$ is much smaller than the one obtained
by keeping the curvature only in the overall factor.

\subsection{Solution at $r\rightarrow0$ (vanishing curvature)\label{sub:Solution-at-r0}}

When we put $r=0$ in the susceptibility (\ref{eq:ChiTildeDimless}),
the gap equation (\ref{eq:inteq-2}) simplifies considerably. First,
the fermionic self-energy loses its dependence on $k_{y}$ and becomes
\begin{equation}
\Sigma(k_{y},\omega_{m})=\Sigma(\omega_{m})=\sgn\omega_{m}\frac{3}{4}\pi T\sum_{|\Omega_{n}|<|\omega_{m}|}|\Omega_{n}|^{-1/4}.\label{eq:SigmaOmega}\end{equation}
 We used the full summation in the numerical calculations below, but
this is actually not even necessary -- we found that for $T=T_{c}$,
the functional form of the self-energy is close to the zero-temperature
expression $\sgn\omega_{m}|\omega_{m}|^{3/4}$ for all Matsubara frequencies.

Second, without the $r$ term in $\tilde{\chi}$, momentum dependence
of the kernel (\ref{10-2}) is of a convolution type: $K(k_{y},k_{y}^{\prime})=K(k_{y}-k_{y}^{\prime})$.
This implies that Eq. (\ref{eq:inteq-2}) is local in real space coordinate
$y$ conjugate to the momentum $k_{y}$, as can be easily seen by
taking reverse Fourier transform in $k_{y}$. The kernel $K(k_{y}-k_{y}^{\prime})$
is not double integrable:\begin{equation}
\int_{-\infty}^{\infty}\left|K(k,k')\right|dkdk'=\infty\end{equation}
 and so the integral equation in $k_{y}$ is non-Fredholmian, which
means it does not have a countable spectrum with integrable eigenfunctions.
However, the kernel is still uniformly finite\begin{equation}
\int_{-\infty}^{\infty}\left|K(k,k')\right|dk=\int_{-\infty}^{\infty}\left|K(k,k')\right|dk'<\infty,\end{equation}
 and thus may have finite solutions. Two such solutions can be easily
guessed: a symmetric one: $\Phi(k_{y},\omega_{n})=\Phi(\omega_{n})$
and an antisymmetric one: $\Phi(k_{y},\omega_{n})=k_{y}\Phi(\omega_{n})$.
As functions of real space coordinate $y$ these two solutions correspond
to $\delta(y)\Phi(\omega_{n})$ and $-(d\delta(y)/dy)\Phi(\omega_{n})$,
respectively, where $\delta(y)$ is a delta function. The $d-$wave
solution we are interested is the one antisymmetric in $k_{y}$. In
both cases, integrating explicitly over momentum and substituting
the result for the susceptibility, we obtain one-dimensional integral
equation for $\Phi(\omega_{n})$: \begin{eqnarray}
 &  & \Phi(\omega_{n})=\frac{3}{4}\pi T\nonumber \\
 &  & \times\sum_{\omega_{m}}\frac{\Phi(\omega_{m})}{|\omega_{n}-\omega_{m}|^{1/4}\left(|\omega_{m}|+\frac{3}{4}\pi T\sum_{|\Omega_{n}|<|\omega_{m}|}|\Omega_{n}|^{-1/4}\right)}\label{eq:integ-gamma}\end{eqnarray}
 Approximating the self-energy by the $\omega^{3/4}$ form simplifies
(\ref{eq:integ-gamma}) to \begin{equation}
\Phi(\omega_{n})=\frac{3}{4}\pi T\sum_{\omega_{m}}\frac{\Phi(\omega_{m})}{|\omega_{m}+\Sigma(\omega_{m})||\omega_{n}-\omega_{m}|^{1/4}}.\label{eq:integ-1/4}\end{equation}
 This equation falls into a generic class of local Eliashberg-type
gap equations with the effective local pairing interaction $\chi_{l}(\omega)\propto\omega^{-\gamma}$
for which the fermionic self-energy scales as $\omega^{1-\gamma}$
(\cite{unpubl}, see also Appendix \ref{sec:Eliashberg-omega}). Eq.
(\ref{eq:integ-1/4}) corresponds to $\gamma=1/4$.

Note that: (i) the gap equation is fully universal and parameter-free
(we recall that $T$ is in units of $\omega_{0}$), and (ii) that
self-energy term is not small and cannot be neglected. At small $\omega_{m}=\pi T(2m+1)<1$,
the kernel of the gap equation scales as $1/\omega$. This power is
a combination of $1/\omega^{1/4}$ from the effective local pairing
interaction and $\omega^{3/4}$ from the fermionic self-energy. At
$\omega_{m}>1$, the kernel decays as $1/\omega^{5/4}$, i.e., faster
than $1/\omega$, and the frequency sum converges even without $\Phi(\omega_{m})$.
Ref. \cite{unpubl} argued that, this gap equation has a non-zero
solution at $T=O(1)$.

We solved this equation numerically (the details are presented in
Appendix \ref{sec:Eliashberg-omega}) and found, in actual units \begin{equation}
T_{c}(r\rightarrow0)\approx6\omega_{0}\equiv6\frac{\overline{g}}{4\pi}r,\label{eq:11-1}\end{equation}
 The scaling with the upper cutoff of the quantum-critical behavior
($\omega_{0}$ in our case) is the same as in hole-doped materials
(where the pairing involves predominantly antinodal fermions), however,
we emphasize that in our case the scale $\omega_{0}$ is by itself
proportional to the Fermi surface curvature, while the corresponding
scale for antinodal pairing remains finite in the absence of the curvature.

Note in passing that the prefactor is much larger in our case than
for antinodal pairing, where $T\approx0.2$ in units of the upper
cutoff of the quantum-critical behavior ~\cite{abanov03}.

\subsection{Solution at finite $r$\label{sub:Solution-at-finite}}

\subsubsection{Numerical solution\label{sub:Numerical-solution}}

Although $T_{c}$ given by (\ref{eq:11-1}) is finite, the solution
of the gap equation without the $r^{2}k_{y}^{4}$ term in the susceptibility
is unphysical since the pairing gap continuously increases with $k_{y}$:
$\Phi(k_{y},\omega_{n})\propto k_{y}$. This behavior reflects translational
invariance of the pairing potential (or, in mathematical language,
the non-Fredholmian property of the integral equation). When the $r^{2}k^{4}$
term is kept in $\widetilde{\chi}(k_{y},k_{y}',\omega)$, Eq. (\ref{eq:inteq-2})
becomes Fredholmian, its solutions integrable, and the gap vanishes
at large $k_{y}$. A numerical solution of the gap equation can then
be obtained by standard technique. We solved Eq. (\ref{eq:inteq-2})
numerically in the quadrant $\omega,k>0$, searching for a solution
$\Phi_{n}(k)$ that is symmetric in $\omega_{n}$ and antisymmetric
in $k$. The semiinfinite integral over $k$ was mapped onto the interval
$[0,1)$ using the transformation $x=(1-k)/k$, and then integration
was approximated by Gauss-Legendre quadrature: $\int_{0}^{1}f(x)dx=\sum_{i}f(x_{i})w_{i}$,
where $x_{i}$ and $w_{i}$are the abscissas and weights of the quadrature.
This procedure is called the Nystrom method \cite{nr}. The integral
equation then reduces to an (infinite in the Matsubara frequency index
$n$) set of algebraic equations: \begin{eqnarray}
\Phi_{ni} & = & \sum_{m\geq0,j}(K_{nimj}+K_{ni,-m-1,j}\nonumber \\
 &  & -K_{nim,-j}-K_{ni,-m-1,-j})\Phi_{mj},\end{eqnarray}
 This set was truncated and solved by standard LAPACK routines. This
procedure gives good approximation to the largest eigenvalues of the
kernel, and for the critical temperature we need just the largest
one.

We found that the effect of the curvature on $T_{c}$ is very strong:
above $r>0.001$, the actual $T_{c}$ rapidly becomes much smaller
than (\ref{eq:11-1}). We plot $T_{c}(r)$ in Fig. \ref{cap:Tc} in
units of $\overline{g}/4\pi$. We see that over a wide range $0.01<r<0.1$,
$T_{c}\approx0.0005\bar{g}$ forms a plateau and weakly depends on
$r$. Using the same $\overline{g}\sim1.6$eV and the $t$-$t^{\prime}$
dispersion as in hole-doped materials~\cite{abanov02} but with positive
chemical potential to reproduce the Fermi surface in Fig. 1, we obtained
$r\sim0.08$, which is within the region where $T_{c}$ is almost
constant, and $T_{c}\sim10$K. For hole-doped cuprates, for the same
parameters, the onset of the pairing was earlier estimated at $T_{c}\sim0.01\overline{g}\sim200K$~\cite{abanov03},
although this number may also be reduced by the curvature of the Fermi
surface.

The theoretical value of $T_{c}\sim10$K at a magnetic QCP in electron-doped
cuprates is in reasonable agreement with the experiment \cite{el-pd,quazilbash05,zimmers05}.
Maximum $T_{c}$ in Nd$_{2-x}$Ce$_{x}$CuO$_{4}$ and Pr$_{2-x}$Ce$_{x}$CuO$_{4}$
is about $20-25$K, but, according to ~\cite{zimmers05}, $T_{c}$
initially increases once the system becomes magnetically ordered.
In any event, even $50\%$ agreement is quite reasonable given the
number of approximations in our theoretical analysis. Note that if
we used Eq. (\ref{eq:11-1}) for $T_{c}$ instead of the correct result,
we would have obtained a very large $T_{c}\sim600$K for the same
parameters. This shows that the Fermi surface curvature truly plays
a major role when the pairing involves near-nodal fermions.

The reduction of $T_{c}$ in electron-doped cuprates, compared to
hole-doped cuprates with the same interaction strength was also obtained
in FLEX calculations \cite{manske00}, although the reported difference
was less drastic than in our analysis.

\begin{figure}
\includegraphics[%
  width=0.99\columnwidth]{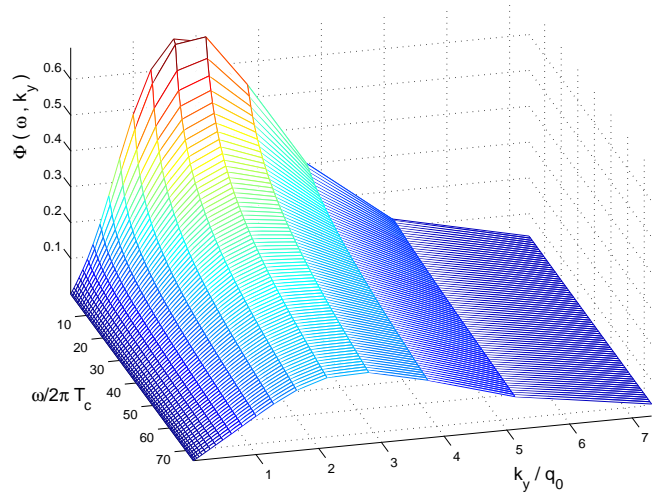}

\caption{The pairing vertex $\Phi(k_{y},\omega_{n})$ for $r=0.05$ vs $\omega_{n}/(2\pi T_{c})$
and $k_{y}/q_{0}$ ($q_{0}$ is defined in the text). Observe that
$\Phi(k_{y},\omega_{n})$ is non-monotonic in $k_{y}$.\label{cap:gap}}
\end{figure}

In Fig \ref{cap:gap} we present the result for the momentum dependence
of the pairing vertex at various frequencies. We see that the pairing
gap is a non-monotonic function of $k_{y}$: it initially increases
with $k_{y}$, but then passes through a maximum and decreases at
larger $k_{y}$. This result agrees with earlier BCS calculations
\cite{yakovenko04}, but we emphasize that our gap equation is of
non-BCS form. We also emphasize that the position of the maximum in
$\Phi(k_{y},\omega)$ is disconnected from the location of the hot
spots which in our QCP analysis are exactly along the Brillouin zone
diagonals, i.e., at $k_{y}=0$.

\subsubsection{Toy model\label{sub:Toy-model}}

The large discrepancy between $T_{c}(r)$ and (\ref{eq:11-1}) can
also be understood analytically, by expanding $T_{c}$ in $r$ beyond
the $O(r)$ term. This expansion is rather non-trivial as one has
to expand around a solution which diverges at large $k_{y}$, and
the divergence must be cut by the corrections which then obviously
are not small at large $k_{y}$. Since this effect is unrelated to
the summation over Matsubara frequencies, which at $T_{c}$ is confined
to first few $m$, we can simplify the model by dropping the frequency
summation, and instead of (\ref{eq:inteq-2}), analyze the approximate
form of the gap equation \begin{equation}
\Phi(k)=\frac{3}{4\pi}\int\frac{\Phi(k')dk'}{(k-k')^{2}+r^{2}(k^{2}+k'^{2})^{2}+x^{2}},\label{11-3}\end{equation}
 In (\ref{11-3}) we set $x^{4}=a(T/\omega_{0})$ with $a\approx3^{3}/2^{9}$
to match Eq. (\ref{eq:11-1}) at vanishing $r$. To shorten notations,
we dropped the subscript $y$ from $k_{y}$.

Equation (\ref{11-3}) can be analyzed analytically. Our goal is to
understand why $T_{c}$ drops by two orders of magnitude compared
to $6\omega_{0}$ already at relatively small $r\sim0.1$.

We search for the solution of (\ref{11-3}) in the form \begin{equation}
\Phi(k)=\Phi_{0}(k)+\delta\Phi(k),\label{11-4}\end{equation}
 where $\Phi_{0}(k)=k$ is the unphysical solution at $r\to0$ found
in Subsection \ref{sub:Solution-at-r0}, and consider $\delta\Phi(k)$
as perturbation. Substituting into (\ref{11-3}) and expanding in
$\delta\Phi$, we obtain \begin{eqnarray}
\delta\Phi(k) & - & \frac{3}{4\pi}\int\frac{\delta\Phi(k')dk'}{(k-k')^{2}+x_{0}^{2}}=\frac{3k}{4\sqrt{x^{2}+4r^{2}k^{4}}}-k\nonumber \\
 & \approx & \left(\frac{3}{4x}-1\right)k-\frac{3r^{2}k^{5}}{2x_{0}^{3}}\label{11-5}\end{eqnarray}
 where $x_{0}=3/4$, and we dropped the regular $O(r)$ terms from
the integral \begin{equation}
\int\frac{\Phi_{0}(k')dk'}{(k-k')^{2}+r^{2}(k^{2}+k'^{2})^{2}+x^{2}}\end{equation}
 and the higher-order terms in the expansion of the square-root in
(\ref{11-5}). We will see below that relevant $k$ are of order $r^{-2/5}$
and $3/4x-1$ is of order $r^{2/5}$, so that $r^{2}k^{4}\sim r^{2/5}$,
which justifies both the expansion of the square-root in (\ref{11-5})
and dropping the regular $O(r)$ terms.

\begin{figure}
\includegraphics[%
  width=1\columnwidth,
  keepaspectratio]{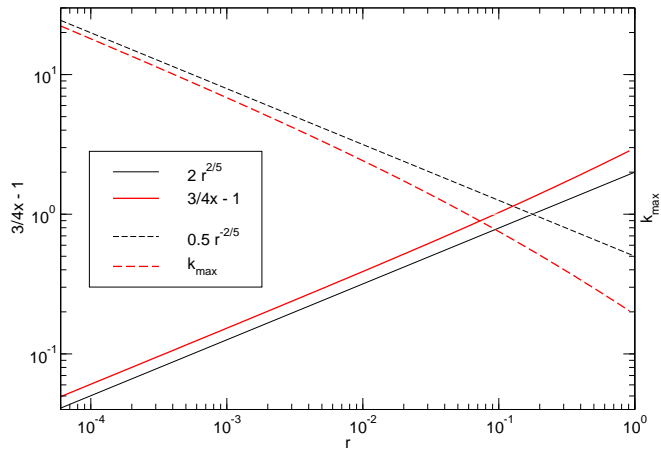}

\caption{Scaling of the {}``critical temperature'' $x(r)$ (solid lines)
and the position of the maximum $k_{\textrm{m}}(r)$ (dashed lines)
with the parameter $r$ in the kernel (\ref{eq:ChiTildeDimless})
of the one-dimensional {}``toy model''. Numerical solution (red
curves) closely follow the scaling laws obtained analytically (black
straight lines): $3/4x-1\approx2r^{2/5}$, and $k_{\mathrm{m}}\sim r^{-2/5}$.\label{cap:Scaling-toyModel}}
\end{figure}

\begin{figure}
\includegraphics[%
  width=1\columnwidth,
  keepaspectratio]{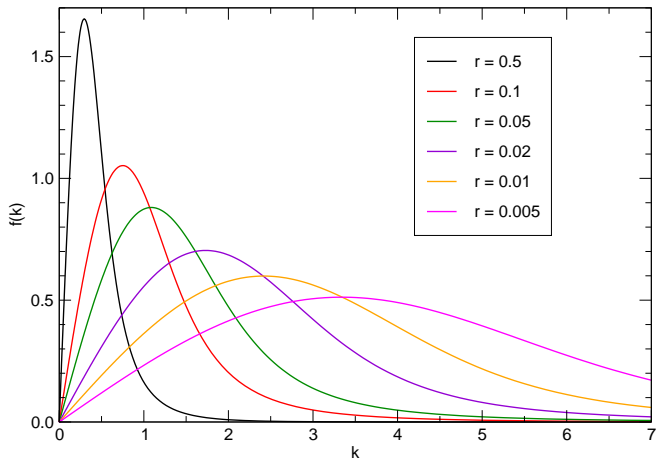}

\caption{Solutions $f(k)$ of the one-dimensional {}``toy model'' for several
values of the parameter $r$ ($r$ decreases for the curves from left
to right). Dependence of the position of the maximum $k_{\mathrm{m}}(r)$
is plotted in Fig. \ref{cap:Scaling-toyModel}. \label{cap:Solutions-toyModel}}
\end{figure}

We search for the solution of (\ref{11-5}) in the form \begin{equation}
\delta\Phi(k)=r^{2}\left(Ak^{3}+Bk^{5}\right),\qquad|k|<k_{\mathrm{m}},\label{11-6}\end{equation}
 where $k_{\mathrm{m}}$ is a cutoff that has to be found from an
auxiliary condition that the solution (\ref{11-4}) with $\delta\Phi(k)$
in the form of (\ref{11-6}) is unique. Substituting (\ref{11-6})
into (\ref{11-5}) and using \begin{eqnarray}
\int_{-k_{\mathrm{m}}}^{k_{\mathrm{m}}}\frac{k'^{3}dk'}{(k-k')^{2}+x_{0}^{2}} & \approx & k^{3}\frac{\pi}{x_{0}}+6kk_{\mathrm{m}}\\
\int_{-k_{\mathrm{m}}}^{k_{\mathrm{m}}}\frac{k'^{5}dk'}{(k-k')^{2}+x_{0}^{2}} & \approx & k^{5}\frac{\pi}{x_{0}}+20k^{3}k_{\mathrm{m}}+\frac{10}{3}kk_{\mathrm{m}}^{3}\end{eqnarray}
 we obtain \begin{eqnarray}
 &  & \frac{3}{4\pi}\int\frac{\delta\Phi(k')dk'}{(k-k')^{2}+x_{0}^{2}}\approx\\
 &  & \frac{3r^{2}}{4\pi}\left[k(6k_{\mathrm{m}}A+\frac{10}{3}k_{\mathrm{m}}^{3}B)+k^{3}(\frac{\pi}{x_{0}}A+20k_{\mathrm{m}}B)+k^{5}\frac{\pi}{x_{0}}B\right].\nonumber \end{eqnarray}
 Substituting this into (\ref{11-5}) and equating the prefactors
for $k$, $k^{3}$ and $k^{5}$ terms in (\ref{11-5}), we obtain
the set of equations\begin{eqnarray}
\left(\frac{3}{4x}-1\right)B & = & \frac{3}{2x_{0}^{3}}\\
\left(\frac{3}{4x}-1\right)A & = & -\frac{15}{\pi}k_{\mathrm{m}}B\label{11-8}\\
-\frac{r^{2}}{2\pi}(9k_{\mathrm{m}}A+5k_{\mathrm{m}}^{3}B) & = & \left(\frac{3}{4x}-1\right)\end{eqnarray}
 Solving the set, we obtain \begin{equation}
\frac{(3/4x-1)^{5}}{r^{2}}=\frac{80u^{2}}{9\pi}\left(\frac{27}{\pi}-u\right),\end{equation}
 where $u=k_{\mathrm{m}}(3/4x-1)$. A unique solution of this equation
exists when $u=18/\pi$ and \begin{equation}
\left(\frac{3}{4x}-1\right)\approx r^{2/5}\left(\frac{320*81}{\pi^{4}}\right)^{1/5}\approx3r^{2/5}\label{11-9}\end{equation}
 Using the definition of $x$ we then obtain \begin{equation}
T_{c}\approx6\frac{\bar{g}}{4\pi}\frac{r}{(1+3r^{2/5})^{4}}\label{11-10}\end{equation}
 This formula is formally valid at small $r$, but it works surprisingly
well for all $r<1$ as evidenced by the comparison of (\ref{11-10})
with the numerical solution of (\ref{11-5}). We compare analytic
and numerical results in Fig. \ref{cap:Scaling-toyModel}, \ref{cap:Solutions-toyModel}.

Eq. (\ref{11-10}) shows that $T_{c}$ does indeed drop quite substantially,
compared to the asymptotic linear behavior (\ref{eq:11-1}): at $r\sim0.1$,
$(1+3r^{2/5})^{4}\approx23$. Furthermore, $T_{c}$ given by (\ref{11-10})
is rather flat at intermediate $r\sim0.04-0.5$ where $T_{c}\sim0.02{\bar{g}}$.
This is consistent with the numerical solution of (\ref{11-5}). The
flat behavior in a wide range of $r$ reproduces what we have found
numerically for the full gap equation (with the full frequency summation).

We also see from (\ref{11-8}) that $A\sim r^{-6/5}$, $B\sim r^{-2/5}$,
and the cutoff $k_{\mathrm{m}}$ diverges as \begin{equation}
k_{\mathrm{m}}=u/\left(\frac{3}{4x}-1\right)\sim r^{-2/5}\label{11-11}\end{equation}
 Substituting $k_{\mathrm{m}}$ and $A$ and $B$ into (\ref{11-6}),
we obtain \begin{equation}
\delta\Phi(k)\sim\Phi_{0}(k)\left(\left(k/k_{\mathrm{m}}\right)^{2}+a\left(k/k_{\mathrm{m}}\right)^{4}\right),\qquad a=O(1)\label{11-14}\end{equation}
 We see that at $k\sim k_{\mathrm{m}}$, the corrections to zero-order
solution become of order 1. A perturbation theory does not allow one
to go beyond this scale (i.e., to analyze $\delta\Phi(k)$ for $k>k_{\mathrm{m}}$),
but it is reasonable to assume that above $k_{\mathrm{m}}$, $\Phi(k)=\Phi_{0}(k)+\delta\Phi(k)$
begins decreasing. The numerical solution of (\ref{11-5}) confirms
this (see Fig. \ref{cap:Solutions-toyModel}). We see from Fig. \ref{cap:Solutions-toyModel}
that the gap in the toy model is a non-monotonic function of $k_{y}$:
it is linear at small $k_{y}$, passes through a maximum at $k_{y}\sim k_{\mathrm{m}}$,
and then falls off. This behavior is again fully consistent with what
we have obtained numerically for the full gap equation.

Returning to the actual, unrescaled momenta $k_{y}$, we obtain, using
the definition of $r$, that the maximum of the gap is located at
\begin{equation}
k_{\mathrm{m}}=\frac{\overline{g}}{\pi v_{F}}k_{\mathrm{m}}\sim\frac{\overline{g}}{v_{F}}r^{-2/5}\sim k_{F}r^{3/5}\left(\frac{\beta_{0}}{\beta}\right)^{2}\label{11-15}\end{equation}
 where $\beta_{0}^{2}=v_{F}/k_{F}$. For $\beta\sim\beta_{0}$, and
$r\ll1$, $k_{\mathrm{m}}$ is much less than $k_{F}$, i.e., the
$d-$wave gap is confined to a near vicinity of the zone diagonal.

We fitted the actual $T_{c}$ for the full model by the functional
form of Eq. (\ref{11-10}), using the prefactor for the $r^{2/5}$
term as the adjusting parameter. We found that the numerical data
are reproduced reasonably well if we set this prefactor to $5$ (instead
of $3$ in Eq. (\ref{11-10})). We show the fit in Fig. \ref{cap:Tc}.

\subsection{Away from the QCP}

At deviations from the QCP towards larger dopings, i.e., into paramagnetic
phase, $T_{c}$ decreases and eventually disappears. The value of
$k_{\mathrm{m}}$, however, does not track the decrease of $T_{c}$,
i.e., the $d-$wave gap extends over a finite momentum range along
the FS even in the overdoped materials. At deviations into antiferromagnetic
phase, the FS evolves into hole and electron pockets, and the locations
of $k_{\mathrm{m}}$ gradually approach the locations of the hot spots.

Our results for $T_{c}$ and the gap survive even when the magnetic
correlation length $\xi$ remains finite at the doping where $2k_{F}=Q$
and the Fermi surface has the form shown in Fig. 1. In this situation,
the antiferromagnetic QCP shifts to lower dopings, when the hot spots
are already away from the Brillouin zone diagonals. We found that
the modifications of our results are small as long as $T_{c}>J\xi^{-2}$
where $J\sim v_{F}^{2}/{\bar{g}}$ is of the order of the the exchange
interaction.

\section{Summary \label{sec:Summary}}

We considered the normal state properties and pairing near a $2k_{F}$
antiferromagnetic QCP and applied the results to electron-doped cuprates
at optimal doping. We found that the breakdown of the Fermi-liquid
description at QCP leads to peculiar frequency dependences of the
conductivity and the $B_{2g}$ Raman response. We found that $T_{c}$
remains finite at the QCP. The pairing gap at QCP has $d_{x^{2}-y^{2}}$
symmetry, but is highly anisotropic and confined to momenta near the
zone diagonals. The value of $T_{c}$ is about $10$K for the same
spin-fermion coupling as for hole-doped cuprates, where the pairing
predominantly involves antinodal fermions. The strong reduction of
$T_{c}$ compared to hole-doped case is due to the fact that for the
$d-$wave pairing of near-nodal fermions, the curvature of the Fermi
surface plays a major role.

We acknowledge useful discussions with G. Blumberg, D. Drew, I. Eremin,
R. Greene, C. Homes, D. Manske, A. Millis, B.S. Mityagin, M. Norman,
S.P. Novikov, M. M. Qazilbash, and V. Yakovenko. The research is supported
by Condensed Matter Theory Center at UMD (P.K, A.C) and by NSF DMR
0240238 (A.C.).

\appendix

\section{Transformation of the polarization operator when hot spots merge
\label{sec:Transformation-of-Pi}}

In this Appendix we detail the transformation of the polarization
operator with the shift in the position of the hot spots along the
Fermi surface. In hole-doped materials the spots with the strong electron
interaction are close to the $(0,\pi)$ and symmetry-related points
on the Fermi surface in the notations of Fig. \ref{cap:sketch}. With
doping and expansion of the Fermi surface its boundary crosses the
antiferromagnetic Brillouin zone at points that become closer to each
other and to the zone diagonal until in electron-doped cuprates they
merge pairwise at the doping level when the Fermi surface just touches
the antiferromagnetic Brillouin zone at $(\pi/2,\pi/2)$ (this situation
is shown in Fig. \ref{cap:sketch}). In hole-doped materials it sufficed
to take linear approximation of electron dispersion near the hot spots:
$\epsilon_{\mathbf{k}}=\mathbf{v}_{F}\mathbf{k}$. The polarization
operator of non-interacting fermions then had the Landau damping form
\begin{equation}
\Pi(\mathbf{q},\Omega)=g^{2}|\Omega|/4\pi v_{x}v_{y},\label{eq:PibareLandau}\end{equation}
 where $g$ is the coupling, and $\mathbf{v}_{F}=(v_{x},v_{y})$ \cite{abanov03}.
In the situation depicted in Fig. \ref{cap:sketch} one has to keep
second order terms in $\epsilon_{\mathbf{k}}$ (\ref{eq:ep}), and
the bare polarization operator has the form (\ref{eq:PiBare}). Here
we show how to go from (\ref{eq:PibareLandau}) to (\ref{eq:PiBare})
with continuous change in the electron spectrum. For later use we
will carry out slightly more general calculation keeping fermionic
self-energy that depends only on frequency. Consider a spectrum that
has both linear term in transverse momentum and curvature:

\begin{eqnarray}
\epsilon_{\mathbf{k}} & = & v_{x}k_{x}+v_{y}k_{y}+\beta^{2}k_{y}^{2},\\
\epsilon_{\mathbf{k}+\mathbf{Q}} & = & -v_{x}k_{x}+v_{y}k_{y}+\beta^{2}k_{y}^{2}.\end{eqnarray}
 The case $\beta=0$ gives dispersion around well-separated hot-spots
of hole-doped materials, and in the case $v_{y}=0$ we revert to (\ref{eq:ep}).
To alleviate integration in $\mathbf{k}$ we note that the following
equality holds\begin{equation}
\epsilon_{\mathbf{k}}+\epsilon_{\mathbf{k}+\mathbf{q}+\mathbf{Q}}=\widetilde{E}_{\mathbf{q}}+y^{2},\end{equation}
 where \begin{equation}
\widetilde{E}_{\mathbf{q}}=-v_{x}q_{x}+\frac{\beta^{2}q_{y}^{2}}{2}-\frac{v_{y}^{2}}{2\beta^{2}}\end{equation}
 differs from $E_{\mathbf{q}}$ (\ref{eq:Eq}) by $-v_{y}^{2}/2\beta^{2}$,
and \begin{equation}
y=\sqrt{2}\beta\left(k_{y}+\frac{q_{y}}{2}+\frac{v_{y}}{2\beta^{2}}\right)\end{equation}
 is a convenient dummy variable in the integration instead of $k_{y}$.
The Jacobian of the substitution $\mathbf{k}\rightarrow(\epsilon_{\mathbf{k}},y)$
is \begin{equation}
\left|\frac{\partial(\epsilon_{\mathbf{k}},y)}{\partial(k_{x},k_{y})}\right|=\sqrt{2}v_{x}\beta.\end{equation}

Momentum integration in (\ref{eq:Pi}) then gives \begin{eqnarray}
 &  & \Pi(\mathbf{q},\Omega)=2g^{2}\int\frac{d\omega d^{2}\mathbf{k}}{(2\pi)^{3}}G\left(\mathbf{k},\omega-\ts\frac{\Omega}{2}\right)G\left(\mathbf{k}+\mathbf{q},\omega+\ts\frac{\Omega}{2}\right)\nonumber \\
 &  & =\frac{ig^{2}}{4\pi\sqrt{2}v_{x}\beta}\int\frac{\sgn\left(\omega+\frac{|\Omega|}{2}\right)+\sgn\left(\omega-\frac{|\Omega|}{2}\right)}{\sqrt{\widetilde{E}_{\mathbf{q}}-i\left(\widetilde{\Sigma}_{\omega+\frac{|\Omega|}{2}}+\widetilde{\Sigma}_{\omega-\frac{|\Omega|}{2}}\right)}}d\omega,\end{eqnarray}
 where $\widetilde{\Sigma}_{\omega}=\omega+\Sigma(\omega)$. Simplifying
frequency integration from the antisymmetric integrand\begin{equation}
\Pi(\mathbf{q},\Omega)=\frac{g^{2}}{\sqrt{2}\pi v_{x}\beta}\int\limits _{|\Omega|/2}^{\infty}\Im\frac{d\omega}{\sqrt{\widetilde{E}_{\mathbf{q}}+i\left(\widetilde{\Sigma}_{\omega+\frac{|\Omega|}{2}}+\widetilde{\Sigma}_{\omega-\frac{|\Omega|}{2}}\right)}},\end{equation}
 or, with the substitute $\omega\rightarrow\omega+|\Omega|/2$, \begin{equation}
\Pi(\mathbf{q},\Omega)=-\frac{g^{2}}{2\pi v_{x}\beta}\int_{0}^{\infty}\frac{\sqrt{\sqrt{\widetilde{E}_{\mathbf{q}}^{2}+(\widetilde{\Sigma}_{\omega}+\widetilde{\Sigma}_{\omega+|\Omega|})^{2}}-\widetilde{E}_{\mathbf{q}}}}{\sqrt{\widetilde{E}_{\mathbf{q}}^{2}+(\widetilde{\Sigma}_{\omega}+\widetilde{\Sigma}_{\omega+|\Omega|})^{2}}}d\omega.\end{equation}
 When $\Sigma(\omega)=0$ the integral above can be taken exactly
and gives \begin{equation}
\Pi(\mathbf{q},\Omega)=\frac{g^{2}}{2\pi v_{x}\beta}\sqrt{\sqrt{\Omega^{2}+\widetilde{E}_{\mathbf{q}}^{2}}+\widetilde{E}_{\mathbf{q}}},\label{eq:PiBareGen}\end{equation}
 which coincides with (\ref{eq:PiBare}) when $v_{y}=0$ and so $\widetilde{E}_{\mathbf{q}}=E_{\mathbf{q}}$.
In the opposite limit $\beta\rightarrow0$, $\widetilde{E}_{\mathbf{q}}\rightarrow-v_{y}^{2}/2\beta^{2}$
and, expanding (\ref{eq:PiBareGen}) up to first order in $|\Omega|/\widetilde{E}_{\mathbf{q}}$
gives \begin{equation}
\Pi(\mathbf{q},\Omega)=g^{2}|\Omega|/2\pi v_{x}v_{y},\end{equation}
 i.e. exactly twice the expression (\ref{eq:PibareLandau}), accounting
for the pair of hot spots that emerge.

\section{Self-energy at $T=0$\label{sec:Self-energy-at-T=3D3D3D0}}

In this Appendix we find self-consistent fermionic self-energy $\Sigma(\omega)$
at low frequencies as explained in Section \ref{sec:Self-consistent-self-energies}.
Following the derivation that leads to Eq. (\ref{eq:PiBare}) it can
be shown that the polarization operator of interacting fermions that
have self-energy $\Sigma(\omega)$ is

\begin{equation}
\Pi(\mathbf{q},\Omega)=-\frac{g^{2}}{2\pi v_{F}\beta}\int_{0}^{\infty}\frac{\sqrt{\sqrt{E_{\mathbf{q}}^{2}+(\widetilde{\Sigma}_{\omega}+\widetilde{\Sigma}_{\omega+|\Omega|})^{2}}-E_{\mathbf{q}}}}{\sqrt{E_{\mathbf{q}}^{2}+(\widetilde{\Sigma}_{\omega}+\widetilde{\Sigma}_{\omega+|\Omega|})^{2}}}d\omega.\end{equation}
 where $\widetilde{\Sigma}_{\omega}=\omega+\Sigma(\omega)$. It is
easy to see that in the case $\Sigma(\omega)=0$ the integral above
can be taken exactly and gives (\ref{eq:PiBare}). Note that we subtract
divergent high-energy contribution \begin{equation}
\Pi_{0}(0,0)=-\frac{g^{2}}{2\pi v_{F}\beta}\int_{0}^{\infty}\frac{d\omega}{\sqrt{2\widetilde{\Sigma}_{\omega}}}\end{equation}
 because it is already included in the theory as the $\xi^{-2}$ {}``mass''
term in the bare susceptibility (\ref{eq:chi0}).

\begin{figure}
\includegraphics[%
  width=0.99\columnwidth]{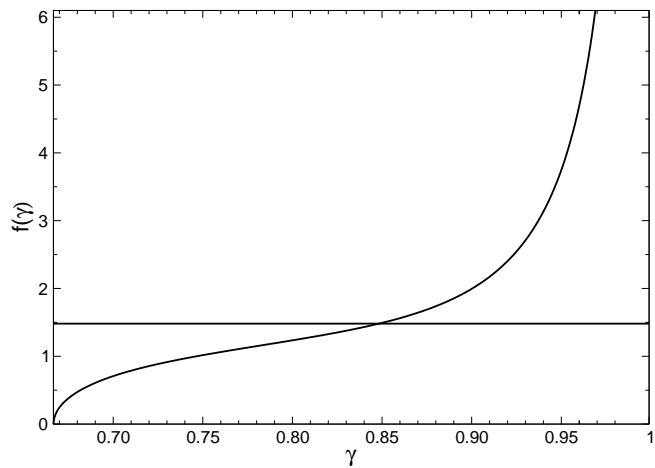}

\caption{Graphical solution to Eq. (\ref{eq:implAlpha}). Straight line and
the curve represent the left and right-hand side of Eq. (\ref{eq:implAlpha}).
Crossing is at about $\gamma\approx0.85$, or $\alpha=1-\gamma\approx0.15$.
\label{cap:Graphical-solution}}
\end{figure}

The above expression for $\Pi$ when substituted as the bosonic self-energy
in the bosonic propagator (\ref{eq:ChiWithPhi0}) in Eq. (\ref{eq:Sigma})
yields a self-consistent equation on $\Sigma$. We will seek a solution
to this equation in the form of a power law: $\widetilde{\Sigma}_{\omega}=\sigma\omega^{\gamma}$,
where $\sigma$ is a constant. Using the substitutions $\Omega=|\omega|t$
so that $\widetilde{\Sigma}_{\Omega}=\widetilde{\Sigma}_{\omega}t^{\gamma}$,
and $\omega'=|\omega|ts$, where $\omega'$ is the dummy variable
in $\Pi$, so that $\widetilde{\Sigma}_{\omega'}+\widetilde{\Sigma}_{\omega'+\Omega}=\widetilde{\Sigma}_{\Omega}(s^{\gamma}+(s+1)^{\gamma})$,
and $q=(2\widetilde{\Sigma}_{\Omega})^{1/2}v/\beta$, we arrive at
a condition on the exponent $\gamma$:\begin{equation}
\frac{\pi N}{3\sqrt{2}}=-\frac{1}{\gamma}\int_{0}^{\infty}f(v;\gamma)dv,\label{eq:implAlpha}\end{equation}
 where \begin{eqnarray}
\frac{1}{f(v;\gamma)} & = & \int_{0}^{\infty}\biggl(\frac{\sqrt{\sqrt{v^{4}+(s^{\gamma}+(s+1)^{\gamma})^{2}}+v^{2}}}{\sqrt{v^{4}+(s^{\gamma}+(s+1)^{\gamma})^{2}}}\label{eq:f(alpha)}\\
 & + & \frac{\sqrt{\sqrt{9v^{4}+(s^{\gamma}+(s+1)^{\gamma})^{2}}-3v^{2}}}{\sqrt{9v^{4}+(s^{\gamma}+(s+1)^{\gamma})^{2}}}-\frac{2}{\sqrt{2s^{\gamma}}}\biggr)ds.\nonumber \end{eqnarray}
 The right-hand side of Eq. (\ref{eq:implAlpha}) is a continuous
function of $\gamma$ in $\frac{2}{3}<\gamma<1$. It monotonically
grows from zero at $\gamma=\frac{2}{3}$ and crosses the constant
$\pi N/3\sqrt{2}$ at $\gamma\approx0.85$ (see Fig. \ref{cap:Graphical-solution}).
Recalling that $\alpha$ used in the main body of the paper is $\alpha=1-\gamma$,
we get $\alpha\approx0.15$ as was ascertained.

The calculation of the right-hand side of Eq. (\ref{eq:implAlpha})
takes some care. Integral (\ref{eq:f(alpha)}) diverges at the upper
limit when $\gamma\rightarrow\frac{2}{3}$. Expanding in large $s$
gives for the residual of (\ref{eq:f(alpha)}) \begin{equation}
\int_{s_{\mathrm{m}}}^{\infty}(\ldots)ds\approx\frac{1}{\sqrt{2}}\left(\frac{v^{2}s_{\mathrm{m}}^{1-3\gamma/2}}{2(1-3\gamma/2)}-s_{\mathrm{m}}^{-\gamma/2}+\ldots\right),\end{equation}
 whence $\int f(v;\gamma)dv\sim\sqrt{1-3\gamma/2}$ for $\gamma$
close to $\frac{2}{3}$. In the opposite limit $\gamma\rightarrow1$
it is the integral over $v$ in (\ref{eq:implAlpha}) that diverges
at the upper limit. At large $v$ the scaling substitution $s=v^{2/\gamma}z$
into (\ref{eq:f(alpha)}) separates an integral over $z$:\begin{equation}
\frac{1}{f(v;\gamma)}\approx C(\gamma)v^{\frac{2}{\gamma}-1},\end{equation}
 so the residual of the integral over $v$ in (\ref{eq:implAlpha})\begin{equation}
\int_{v_{\mathrm{m}}}^{\infty}f(v;\gamma)dv\approx-\frac{v_{\mathrm{m}}^{\frac{2}{\gamma}-2}}{(\frac{2}{\gamma}-2)C(\gamma)}\end{equation}
 diverges as $\sim(1-\gamma)^{-1}$. The auxiliary integral\begin{eqnarray}
C(\gamma) & = & \int_{0}^{\infty}\biggl(\frac{\sqrt{\sqrt{1+(2z^{\gamma})^{2}}+1}}{\sqrt{1+(2z^{\gamma})^{2}}}\nonumber \\
 & + & \frac{\sqrt{\sqrt{9+(2z^{\gamma})^{2}}-3}}{\sqrt{9+(2z^{\gamma})^{2}}}-\frac{2}{\sqrt{2z^{\gamma}}}\biggr)dz.\end{eqnarray}
 It is negative in the interval $\frac{2}{3}<\gamma<1$ and diverges
when $\gamma$ approaches $\frac{2}{3}$. When $\gamma=1$ the integral
can be taken analytically: $C(1)=-\sqrt{6}$. To produce Fig. \ref{cap:Graphical-solution}
we combined the numerical integration upto the finite cutoff with
the analytical residuals found above.

\section{Choice of the cutoff for the conductivity\label{sec:Choice-of-cutoff}}

In this Appendix we substantiate our claim that the frequency dependence
of the conductivity does not rely on how the integration over transverse
momenta in (\ref{eq:sM}) is estimated. To evaluate the integral over
$k_{y}$ we assumed that $\Sigma(\epsilon,k_{y})$ stays almost constant
and equal to $\Sigma(\epsilon)$ for $|k_{y}|<k_{y,\mathrm{max}}$
and then falls off rapidly. Crudely we may just replace $\int dk_{y}$
with $k_{y,\mathrm{max}}$. However, since $k_{y,\mathrm{max}}\propto\omega^{\gamma}$
depends on frequency, it is not clear whether we should choose the
external $\omega$ or the internal frequency $\epsilon$ for the cutoff.
Choosing the first one yields\begin{equation}
\sigma(i\omega)\propto\omega^{\gamma-1}\int_{0}^{\omega}\frac{d\epsilon}{\omega+\Sigma(\omega-\epsilon)+\Sigma(\epsilon)}.\end{equation}
 On the other hand, if we simply integrate over $k_{y}$ in (\ref{eq:sM})
with $\Sigma(\epsilon,k_{y})=\Sigma(\epsilon)$ for $|k_{y}|<k_{y,\mathrm{max}}$,
the result is\begin{eqnarray}
\sigma(i\omega) & \propto & 2\int_{0}^{\omega/2}\biggl(\frac{\epsilon^{\gamma}}{\omega+\Sigma(\omega-\epsilon)+\Sigma(\epsilon)}\nonumber \\
 &  & +\frac{(\omega-\epsilon)^{\gamma}-\epsilon^{\gamma}}{\omega+\Sigma(\omega-\epsilon)}\biggr)d\epsilon.\end{eqnarray}
 Intuitively, the choice of the cutoff of the momentum integration
should not alter the frequency dependence of the conductivity since
the internal $\epsilon$ stays of order of the external $\omega$
anyway. And indeed we found that all three estimates lead to the same
pseudo-scaling of $\sigma(\omega)$ in the range $\omega_{0}<\omega<40\omega_{0}$.
The choice of the cutoff alters only the irrelevant preexponent shared
by both $\Re\sigma$ and $\Im\sigma$.

The fact that $\Re\sigma$ and $\Im\sigma$ scale together with $|\sigma(\omega)|$
can be seen from $\arctan(\Im\sigma/\Re\sigma)$ staying almost constant
over the range $\omega_{0}<\omega<40\omega_{0}$ (see Fig. \ref{cap:conductivity},
right panel). The value of the constant $\varphi=\arctan(\Im\sigma/\Re\sigma)$
itself changes with the Wick transition from Matsubara to real frequencies.
In particular, $\varphi$ depends on whether the Matsubara or real
frequency enters the momentum cutoff $k_{y,\mathrm{max}}(\omega)$.
We argue that it is logical to cut off the momentum integration on
Matsubara frequencies since the subsequent frequency integration is
done over Matsubara frequencies as well. With this choice the constant
$\varphi\approx1$ (see Fig. \ref{cap:conductivity}, right panel).

\section{Eliashberg local pairing \label{sec:Eliashberg-omega}}

In this Appendix we describe the numerical solution of the pairing
problem (\ref{eq:integ-gamma}) for a general power-law local interaction.
The results of this Section were used for $T_{c}$ of (\ref{eq:integ-1/4}).

The general interaction is (the frequencies and the temperature are
measured in suitable units which we put unity)\begin{equation}
\chi(\Omega_{n})=\pi(1-\gamma)|\Omega_{n}|^{-\gamma}.\label{eq:Chi-local}\end{equation}
 Fermionic self-energy $\Sigma(\omega_{n})$ corresponding to (\ref{eq:Chi-local})
is given by \begin{equation}
\Sigma(\omega_{n})=T\sum_{\omega_{m}}\sgn\omega_{m}\chi(\omega_{n}-\omega_{m}).\end{equation}
 On substituting $\Omega_{n}=\omega_{n}-\omega_{m}$ the sum in $\Sigma_{n}$
becomes finite:\begin{equation}
\Sigma(\omega_{n})=\sgn\omega_{n}T\sum_{|\Omega_{m}|<|\omega_{n}|}\chi(\Omega_{m}).\label{eq:Sigma-local}\end{equation}
 Approximating the sum by an integral we get the limiting scaling
of the self-energy: \begin{equation}
\Sigma(\omega)\approx\sgn\omega\frac{1}{\pi}\int_{0}^{\omega}\chi(\Omega)d\Omega\equiv\sgn\omega|\omega|^{1-\gamma}\end{equation}

Reverting to the pairing problem and denoting $\widetilde{\Sigma}_{n}=\omega_{n}+\Sigma(\omega_{n})$
and $\Phi_{n}=\Phi(\omega_{n})$, we get for the linearized system
on the self-energy and anomalous vertex at finite temperatures by
analogy to Eqs. (\ref{eq:SigmaOmega}), (\ref{eq:integ-1/4}):\begin{eqnarray}
\widetilde{\Sigma}_{n} & = & \omega_{n}+T\sum_{m}\sgn\omega_{m}\chi(\omega_{n}-\omega_{m})\\
\Phi_{n} & = & T\sum_{m}\frac{\Phi_{m}}{|\widetilde{\Sigma}_{m}|}\chi(\omega_{n}-\omega_{m})\end{eqnarray}

Introducing $\widetilde{\Sigma}_{n}=\omega_{n}Z_{n}$, $\Phi_{n}=\Delta_{n}Z_{n}$
we get one equation on $\Delta_{n}$:\begin{equation}
\Delta_{n}=T\sum_{m}\left(\frac{\Delta_{m}}{\omega_{m}}-\frac{\Delta_{n}}{\omega_{n}}\right)\sgn\omega_{m}\chi(\omega_{n}-\omega_{m})\end{equation}
 Here the summation clearly can be restricted to $m\ne n$. Returning
back to the original formulas we see that we can redefine $\widetilde{\Sigma}_{n}$,
$\Phi_{n}$ as sums with $m\ne n$. (For the self-energy in the form
(\ref{eq:Sigma-local}) this means $\Omega_{n}\ne0$).

Expressing Matsubara frequencies in overt form, bosonic as $\Omega_{n}=2\pi Tn$,
fermionic as $\omega_{n}=2\pi T(n+1/2)$, we may rewrite the finite
sum in the self-energy (\ref{eq:Sigma-local}) via the harmonic numbers
\begin{equation}
H_{n}^{(\gamma)}=\sum_{m=1}^{n}\frac{1}{m^{\gamma}}\end{equation}
 as \begin{equation}
\Sigma(\omega_{n}>0)=(1-\gamma)(2\pi T)^{1-\gamma}H_{n}^{(\gamma)}.\end{equation}

Eq. () on the anomalous vertex $\Phi$ should be understood in the
sense that at some $T_{c}$, which would be the critical temperature
for the onset of pairing, a non-trivial $\Phi(\omega_{n})$ would
become a solution, i.e. at $T_{c}$ the largest eigenvalue of () crosses
unity. The kernel of Eq. () symmetric with respect to simultaneous
change of sign before both $\omega_{n}$ and $\omega_{m}$, so its
solutions must be either even or odd in frequency. From a known theorem,
the largest eigenvalue corresponds to an even eigenvector $\Phi(\omega_{n})=\Phi(-\omega_{n})$.
For the even solution we get the equation\begin{equation}
\Phi_{n}=\frac{1}{2}\sum_{m}'\frac{|n-m|^{-\gamma}+|n+m|^{-\gamma}}{\frac{(2\pi T)^{\gamma}}{1-\gamma}(m+\frac{1}{2})+H_{m}^{(\gamma)}}\Phi_{m}\end{equation}
 where in the term $m=n$ expression $|n-m|^{-\gamma}$ should be
substituted with zero.

\begin{figure}
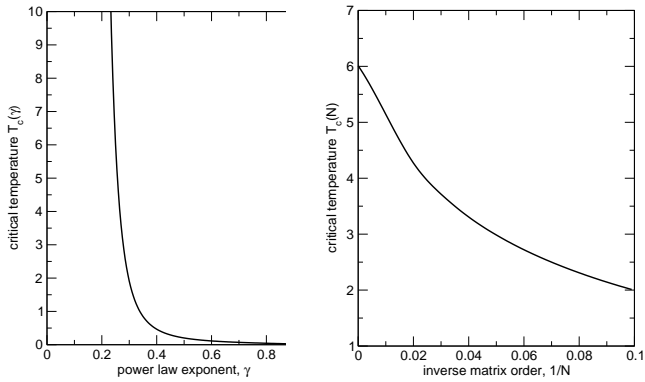

\includegraphics[%
  width=0.49\columnwidth,
  keepaspectratio]{graphics/Tc.eps}\includegraphics[%
  width=0.49\columnwidth]{graphics/TcN.eps}

\caption{Critical temperatures $T_{c}$ found numerically vs. the power law
exponent $\gamma$ (left panel). Right panel represents the dependence
of $T_{c}$ on the matrix size $N$ extrapolated to $1/N\rightarrow0$
for $\gamma=1/4$; behavior for other $\gamma$ is similar; left panel
shows $T_{c}$ already extrapolated to $1/N\rightarrow0$.\label{cap:Critical-temperatures-T}\}}
\end{figure}

The numerical solution of the eigenvalue problem was done with LAPACK
for a finite matrix of order $N$ with several increasing values of
$N$. The sequence of the critical temperatures $T_{c}(N)$ found
for each $N$ is then extrapolated to $1/N\rightarrow0$. A sequence
for $\gamma=1/4$ plotted in Fig. \ref{cap:Critical-temperatures-T}\},
right panel, represents a typical picture. The critical temperature
as function of $\gamma$ are shown in left panel of Fig. \ref{cap:Critical-temperatures-T}\}.

\begin{thebibliography}{15}
\bibitem{damascelli}A. Damascelli, Z. Hussain, and Z.X. Shen, Rev. Mod. Phys. \textbf{75},
473 (2003). 
\bibitem{blumberg02}G. Blumberg, A. Koitzsch, A. Gozar, B. S. Dennis, C. A. Kendziora,
P. Fournier, and R. L. Greene, Phys. Rev. Lett. \textbf{88}, 107002
(2002); ibid. \textbf{90}, 149702 (2003). 
\bibitem{basov}S.V. Dordevic and D.N. Basov, cond-mat/0510351 (unpublished) and references
therein. 
\bibitem{onose04}Y. Onose, Y. Taguchi, K. Ishizaka, and Y. Tokura, Phys. Rev. B \textbf{69},
024504 (2004). 
\bibitem{millis04}A. J. Millis, A. Zimmers, R. P. S. M. Lobo, N. Bontemps, and C. C.
Homes, Phys. Rev. B \textbf{72}, 224517 (2005). 
\bibitem{marel}D. van der Marel, H. J. A. Molegraaf, J. Zaanen, Z. Nussinov, F. Carbone,
A. Damascelli, H. Eisaki, M. Greven, P. H. Kes, M. Li, Nature \textbf{425},
271 (2003). 
\bibitem{homes}C. Homes, unpublished 
\bibitem{camp}Z-X Shen, D. Dessau et al, \prl {\textbf{70}}, 1533 (1993); H. Ding,
\prb {\textbf{54}}, R9678 (1996). 
\bibitem{sato}H. Matsui, K. Terashima, T. Sato, T. Takahashi, M. Fujita, and K.
Yamada, Phys. Rev. Lett. \textbf{95}, 017003 (2005). 
\bibitem{Aiff03}L. Alff, Y. Krockenberger, B. Welter, M. Schonecke, R. Gross, D. Manske,
M. Naito, Nature \textbf{422}, 698 (2003). 
\bibitem{el-pd}Y Tokura, H. Takagi, S. Uchida , Nature \textbf{337}, 345 (1989). 
\bibitem{koitzsch03}A. Koitzsch, G. Blumberg, A. Gozar, B. S. Dennis, P. Fournier, and
R. L. Greene, \prb \textbf{67}, 184522 (2003). 
\bibitem{zimmers05}A. Zimmers, J. M. Tomczak, R. P. S. M. Lobo, N. Bontemps, C. P. Hill,
M. C. Barr, Y. Dagan, R. L. Greene, A. J. Millis and C. C. Homes,
Europhys. Lett. \textbf{70}, 225 (2005). 
\bibitem{piers}P. Coleman and C. Pepin, Acta Physica Polonica B \textbf{34}, 691
(2003). 
\bibitem{dagan04}Y. Dagan, M. M. Qazilbash, C. P. Hill, V. N. Kulkarni, and R. L. Greene,
Phys. Rev. Lett. \textbf{92}, 167001 (2004). 
\bibitem{pines-scalapino}D.J. Scalapino, Phys. Rep. \textbf{250}, 329 (1995); P. Monthoux and
D. Pines, Phys. Rev. B \textbf{47}, 6069 (1993). 
\bibitem{manske00}D. Manske, I. Eremin, and K.H. Bennemann, Phys. Rev. B \textbf{62},
13922 (2000). 
\bibitem{abanov03}A. Abanov, A. V. Chubukov, and J. Schmalian, Adv. Phys. \textbf{52},
119 (2003). 
\bibitem{finn}Ar. Abanov, A. V. Chubukov and A. M. Finkel'stein, Europhys. Lett.,
\textbf{54}, 488 (1999) 
\bibitem{markiewicz03}R.S. Markiewicz, in \emph{Intrinsic Multiscale Structure and Dynamics
in Complex Electronic Oxides}, ed. by A.R. Bishop, et al., World Scientific
(2003). 
\bibitem{onufrieva04}F. Onufrieva and P. Pfeuty, Phys. Rev. Lett. \textbf{92}, 247003 (2004). 
\bibitem{altshuler95}B.L. Altshuler, L.B. Ioffe, A.J. Millis, Phys. Rev. B \textbf{52},
5563 (1995). 
\bibitem{yakovenko04}V. A. Khodel, V. M. Yakovenko, M. V. Zverev, and H. Kang, Phys. Rev.
B \textbf{69}, 144501 (2004). 
\bibitem{num}K. Yoshimura and D.S. Hirashima, J. Phys. Soc. Jpn. \textbf{74}, 712
(2005) and references therein. 
\bibitem{krotkov06}P. Krotkov and A.V. Chubukov, Phys. Rev. Lett. \textbf{96}, 107002
(2006). 
\bibitem{nested}A. Virosztek and J. Ruvalds, Phys. Rev. B \textbf{42}, 4064 (1990). 
\bibitem{comment}The self-energy at finite temperature has been obtained by O. Tchernyshyov
and A.V. Chubukov (unpublished). 
\bibitem{aim}B. L. Altshuler, L. B. Ioffe, and A. J. Millis, singular fermions. 
\bibitem{pepin}A.V. Chubukov, C. Pepin, and J. Rech, Phys. Rev. Lett. \textbf{92},
147003 (2004). 
\bibitem{hlubina95}R. Hlubina, T. M. Rice, \prb \textbf{51}, 9253 (1995). 
\bibitem{norman06}M.R. Norman and A.V. Chubukov, Phys. Rev. B \textbf{73}, 140501 (R)
(2006). 
\bibitem{quazilbash05}M. M. Qazilbash, A. Koitzsch, B. S. Dennis, A. Gozar, Hamza Balci,
C. A. Kendziora, R. L. Greene, G. Blumberg, cond-mat/0510098, to be
published in PRB. 
\bibitem{comm4}A symmetric in $k_{y}$ solution of (\protect\ref{eq:inteq-1}) that
would correspond to a $p-$wave gap stipulates a different overall
sign before the kernel; and an odd-frequency $d-$wave solution has
lower $T_{c}$ than the even-frequency one. 
\bibitem{unpubl}Ar. Abanov, B.L. Altshuler, A.V. Chubukov, and J. Schmalian, (unpublished). 
\bibitem{nr}W. H. Press, B. P. Flannery, S. A. Teukolsky, W. T. Vetterling, \emph{Numerical
Recipes: The Art of Scientific Computing}, New York, 2002. 
\bibitem{abanov02}Ar. Abanov, A. V. Chubukov, M. Eschrig, M. R. Norman, and J. Schmalian,
Phys. Rev. Lett, \textbf{89}, 177002 (2002). 
\end{thebibliography}
\end{document}